# Atomically precise triple-step staircase on a vicinal silicon surface: Is it Si(5 5 7), Si(7 7 10) or Si(8 8 11)?

A. N. Chaika[1], A. Yu. Aladyshkin[2–4], V. N. Semenov[1], A. S. Aladyshkina[5], A. M. Ionov[1], S. I. Bozhko[1]

[1]Osipyan Institute of Solid State Physics Russian Academy of Sciences, Acad. Osipyan str. 2, 142432 Chernogolovka, Russia
[2]Institute for Physics of Microstructures Russian Academy of Sciences, GSP-105, 603950 Nizhny Novgorod, Russia
[3]Center for Advanced Mesoscience and Nanotechnology, Moscow Institute of Physics and Technology, Institutskiy str. 9, 141700 Dolgoprudny, Russia
[4]Lobachevsky State University of Nizhny Novgorod, Gagarin Av. 23, 603022 Nizhny Novgorod, Russia
[5]National Research University Higher School of Economics (HSE University), 603155 Nizhny Novgorod, Russia

## Abstract

Scanning tunneling microscopy studies of periodic arrays of triple steps fabricated on single-crystalline Si(5 5 7) wafers demonstrate several possible atomic structures of consecutive steps and Si(1 1 1) terraces maintaining the same periodicity on micrometer-sized surface areas. Detailed analysis of the atomicallyresolved data reveals the formation of Si(8 8 11) triple-step staircase with a period of $18b \approx 5.99$ nm in projection onto the terrace plane, where $b$=0.333 nm is the distance between atomic rows for the Si(1 1 1)1×1 surface. Schematic models for several possible configurations of either 7×7 or 5×5-reconstructed terraces and triple steps are proposed.

## Introduction

High-Miller-index (or vicinal) Si($h$ $h$ $m$) surfaces consisting of arrays of flat Si(1 1 1) terraces and mono- and multiatomic steps can be considered as atomically precise length standards [1,2], and as templates for the growth of one- and two-dimensional structures on semiconductor substrates that exhibit unique electronic [3-6], transport [7, 8], magnetic [9, 10], and optical [11,12] properties. Kirakosian et al. [1] reported for the first time the fabrication of an atomically precise periodic staircase using Si(5 5 7) wafers. This structure consists of triple steps with a height of $3d$=0.94 nm [13,14], separated by 7×7-reconstructed Si(1 1 1) terraces. Here, $d$=0.3135 nm is the distance between neighboring silicon atomic (bi-)layers in the <111> crystallographic direction. A periodic arrangement of Si(1 1 1) terraces, containing half of the 7×7 unit cell, and triple steps was achieved over surface areas with lateral dimensions up to 200 nm. In the original scanning tunneling microscopy (STM) study the periodicity of the triple-step staircase [1] was determined to be 5.73 nm and the triple step facets were suggested to have a (1 1 2) orientation. The schematic model of the proposed Si(5 5 7) triple step structure providing the period $L^* \approx 5.73$ nm along the vicinal surface is shown in Fig. 1a. This structure corresponds to the period $L = 17b \approx 5.65$ nm in projection onto the Si(1 1 1) terrace plane, where $b$=0.333 nm is the distance between neighboring <110>-directed zigzag atomic rows on the Si(1 1 1)1×1 surface. However, the precise atomic structure of the staircase could not be unambiguously identified due to the complexity of the highly corrugated surface. Several studies employing low-energy electron diffraction (LEED) [15], STM [16–23], atomic force microscopy (AFM) [24, 25], grazing-incidence X-ray diffraction (GIXRD) [26], and density functional theory (DFT) [14] have been conducted to elucidate the structure of the atomically accurate triple-step staircase. These works proposed several alternative models for the periodic Si(*h h m*) triple-step structure. LEED studies by M. Henzler and R. Zhachuk [15] suggest the formation of the Si(5 5 7) staircase with (1 1 3)-oriented triple steps, while STM and DFT studies [17] consider the model with (1 1 2)-oriented triple steps schematically shown in Fig. 1a. In the STM study by Teys et al. [16], the formation of a regular Si(7 7 10) staircase with 7×7-reconstructed terraces and triple steps consisting of a monatomic step, a 3 2/3-atomic-row-wide (111) terrace, and a (001)-oriented double step (Fig. 1b) was proposed. The authors [16] argued that this Si(7 7 10) surface structure with $L = 16b \approx 5.33$ nm (Fig. 1b) was responsible for the formation of the atomically precise step array, while the Si(5 5 7) triple-step staircase [1, 15, 17] could be related to an erroneous interpretation of the experimental data. The model of the triple step facet consisting of consecutive monatomic and double steps separated by a narrow Si(1 1 1) terrace was supported by later studies of vicinal Si(*h h m*) surfaces [19–22, 24–26], although a (1 1 3) orientation of the double steps was proposed in LEED and atomically resolved STM studies [21].

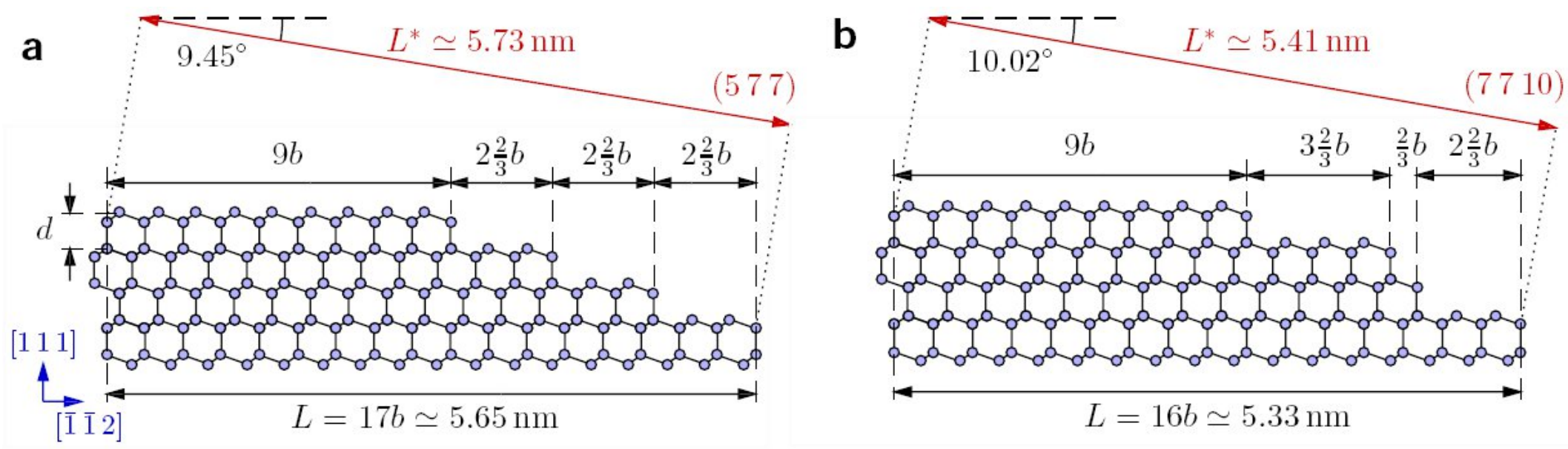

**Figure 1.** Schematic models of the periodic Si(5 5 7) (a) and Si(7 7 10) (b) triple-step staircases with a (112)-oriented triple step (a) and a sequence of monatomic step, Si(1 1 1) mini-terrace and (001)-oriented double step with height equal to two distances 2*d* between neighboring silicon atomic layers (b) [1, 16]. The model presentations do not show adatoms on the 7×7-reconstructed Si(111) terraces and reconstructions at the step edges. The step periodicity in projection onto the terrace plane and along the vicinal surface are indicated as *L* and $L^*$, respectively.

Note that formation of the regular Si(7 7 10) triple-step staircase corresponding to the 10.02° miscut from the (1 1 1) plane was suggested during the STM studies performed on nominally Si(5 5 7) single crystalline wafers with a 9.45° miscut angle [16]. In our previous studies we demonstrated that different interpretations of the triple-step staircases fabricated on the Si(5 5 7) wafers could be related to the difficulties of the precise atomic structure determination on nonflat surfaces [27, 28] and formation of stepped structures with different atomic structures and periodicities on the silicon wafers with the same miscut angle [20--22]. For example, monatomic step array [20] as well as Si(5 5 6) [20] and Si(2 2 3) triple-step staircases [21] were observed on Si(5 5 7) wafers prepared in ultra-high vacuum (UHV) using different thermal treatment procedures. Such strong deflections of the local surface orientation from the macroscopic wafer miscut angle could be related to influence of atomic defects, adsorbates and doping atoms on the terraces and steps [29--31], as well as step-step interaction [32]. Since silicon surface preparation procedures typically involve direct current heating, one should also take into account electromigration phenomena, which could modify the periodicity of the step array and step structure during heating. However, even if an atomically precise, triple-step array with the same atomic structure and periodicity is fabricated on vicinal silicon substrates, different interpretations of the scanning probe microscopy data cannot be excluded. For the proposed models of the triple steps (Fig. 1), the width of the 7×7-reconstructed terrace is equal to $9b$ (half of the 7×7 unit cell and additional adatom row) and comparable to the width of the $3d$ step. In SPM experiments, both the step and terrace facets are inclined at large angles relative to one another and relative to the scanning plane, which significantly complicates the precise analysis of the measured images. The lateral dimensions in STM experiments could also be distorted because of uncontrollable drift and creep effects.

In this work, we utilized an optimized surface preparation procedure in order to fabricate a periodic, regularly spaced triple-step array covering surface areas of at least 0.5 × 0.5 $\mu m^2$. A series of atomically resolved STM studies were carried out on this regular staircase to determine the periodicity and atomic structure of the steps and terraces. LEED patterns, atomically resolved STM images, and differential $D(x,y)$ maps obtained using the difference-of-Gaussians (DOG) approach [27, 28] and corrected using the 7×7 and 5×5 unit cells were analyzed and compared with the results of 2D Fourier map calculations performed for one of the experimentally observed triple step configurations [33]. The periodicity of the triple-step staircase derived from the corrected $D(x,y)$ maps in our case definitively corresponds to $L = 18b \approx 5.99$ nm, corresponding to the Si(8 8 11) surface orientation with an 8.93° miscut angle, which is in contrast to the Si(5 5 7) and Si(7 7 10) triple-step structures discussed previously. The atomically resolved STM data reveal the triple step atomic structure consisting of consecutive monatomic and double steps separated by narrow Si(1 1 1) mini-terraces with fragments of the 7×7 and 5×5 reconstructions. Nevertheless, a unique model of the triple step configuration responsible for the periodic staircase on the vicinal silicon surface can hardly be established. Moreover, our atomically-resolved STM data show that several different configurations of triple steps and either 5×5- or 7×7-reconstructed Si(1 1 1) terraces of different widths can maintain the same periodicity corresponding to the Si(8 8 11) staircase. STM data also reveal the presence of a large number of atomic defects at the steps, which may also contribute to the minimization of the surface energy and stabilization of regularly spaced staircase on micrometer-scale surface areas.

**Methods**

The fabrication of the regular triple-step array on the vicinal silicon wafers, as well as the subsequent LEED and STM experiments, was carried out in the UHV chamber of a LAS-3000 (RIBER) electron spectrometer equipped with a room-temperature scanning tunneling microscope GPI-300 (manufactured by Σ-Scan), where the base pressure remained below $1\times10^{-10}$ mbar. For the STM experiments, [001]-oriented single-crystal tungsten probes were prepared by electrochemical etching of 0.5 × 0.5 × 10 $mm^3$ bars in a 2 M NaOH solution. To obtain a clean and stable tip apex for atomic-resolution experiments, we employed high-temperature electron beam heating and coaxial ion sputtering in the UHV STM chamber [34, 35]. STM experiments were performed at room temperature under constant tunneling current *I* and constant sample bias voltage *U* with respect to a virtually grounded tip. Topography images and differential $D(x,y)$ maps were processed using the freely available WSxM software [36] and our own code written in Matlab. Vicinal silicon wafers (0.5×3×8 $mm^3$, *n*-type, phosphorus-doped, $\rho$ = 25 Ohm·cm at 300 K) for the experiments were cut from single-crystalline ingots with a miscut angle of 9.45° from the Si(1 1 1) plane towards the [-1 -1 2] direction and oriented using X-ray diffraction [37]. The Si(5 5 7) wafers were polished and coated with a protective oxide layer. To create a periodic array of the triple steps with the local surface orientation as close as possible to the macroscopic miscut angle, the Si(5 5 7) wafers were subjected to direct current resistive heating using a special procedure minimizing mass transport and step bunching. The pressure in the UHV chamber during sample heating did not exceed $1\times10^{-9}$ mbar and recovered rapidly to the base values after switching off sample heating. The temperature of the samples was controlled using an optical pyrometer.

**Fabrication of the periodic triple-step array on the Si(5 5 7) wafer**

In our previous work [20], we observed that a periodic array of monatomic steps separated by 7×7-reconstructed Si(1 1 1) terraces containing half of the 7×7 unit cell can be fabricated on the Si(5 5 7) wafers after flash heating to temperatures in the range from 1250°C to 1300°C, rapid cooling to 1050°C, and quenching to temperatures well below the 1×1 → 7×7 phase transition (870°C) with electric current directed either parallel or perpendicular to the step edges. Step bunching and electromigration phenomena did not play a substantial role on the Si(5 5 7) wafers either at the high temperatures used for flashing or at the moderate temperatures used for postannealing. The conversion of monatomic steps to triple steps [13] was not observed in this case. Therefore, we believe that flash heating can be used to remove adsorbates and obtain a uniform distribution of monatomic steps on the vicinal surface. Then, during slower cooling near the phase transition temperature, these monatomic steps on the Si(5 5 7) wafers are expected to convert into triple steps [13]. We also found [20] that continuous heating of the Si(5 5 7) wafers at temperatures close to 870°C can provoke undesirable growth of flat 7×7-reconstructed Si(1 1 1) terraces and stabilization of the Si(5 5 6) triple-step staircase with a substantially larger period [20]. The formation of wide 7×7-reconstructed Si(1 1 1) terraces seems to be energetically favorable when the temperature is close to the phase transition and the silicon surface atoms are sufficiently mobile. The repulsive step–step interaction [32] cannot stabilize regularly spaced narrow terraces with a width corresponding to approximately one 7×7 unit cell or prevent flat terrace growth in this case. Therefore, continuous annealing of the Si(5 5 7) wafers near the phase transition point is undesirable. Cooling from 1050°C to 600°C cannot be too slow if one aims to fabricate a regular Si(*h h m*) staircase with minimal deviation of the local surface orientation from the original miscut angle. One should also take into account that protective oxide layer can be evaporated, and wide 7×7-reconstructed surface areas can be formed on vicinal silicon wafers during a continuous outgassing at elevated temperatures. In this case, a regular triple-step staircase of the desired periodicity can hardly be fabricated on Si(*h h m*) wafers due to the undesirable growth of flat Si(1 1 1)-7×7 areas and step bunching prior to flash heating.

Taking into account the aforementioned observations, the preparation of the triple-step staircase on the Si(5 5 7) wafers in our experiments included the following main steps:

i) continuous outgassing at a moderate temperature of about 600°C for 15–20 hours to remove contamination from the wafers, contact plates, and holder, which allowed us to avoid damaging the protective oxide layer on the sample surface before initiating regular step array formation;

ii) flash heating to 1250–1300°C (several seconds) to remove the protective oxide layer and adsorbates from the surface;

iii) rapid (20–30 seconds) cooling to 900–950°C to avoid step bunching and undesirable growth of Si(1 1 1) terraces, which are possible at intermediate temperatures on vicinal Si(1 1 1) surfaces [38–43];

iv) annealing for approximately 1 minute at 900–950°C to stabilize a uniform step structure on the surface;

v) slow, gradual (15–20 minutes) cooling from 900°C to 600–650°C to form the 7×7 reconstruction on the terraces and to convert monatomic steps to triple steps during the surface phase transition;

vi) continuous (20 minutes to 1 hour) postannealing at 600–650°C.

STM studies of the Si(5 5 7) wafers prepared using different annealing strategies showed the importance of this last stage, which allowed us to complete sample heating under the best vacuum conditions ($1\times10^{-10}$ mbar) and achieve greater regularity of the triple-step array. At higher postannealing temperatures, periodic triple-step systems on large (micron-sized) surface areas were not observed, possibly because of the growth of flat (1 1 1) surface areas. The current was directed perpendicular to the step edges in the step-up direction, as it provided better reproducibility and higher regularity of the triple-step arrays in our experiments.

The application of the above described procedure allowed us to reproducibly fabricate atomically precise, periodic triple-step arrays on surface areas with lateral dimensions exceeding 0.5 µm. As an example, a $D(x,y)$ map obtained from a large-area STM image showing more than 100 periods of the triple-step structure is shown in Fig. 2 (top panel). As can be seen from the $y$-averaged cross-section of the $D(x,y)$ map (bottom panel in Fig. 2), the periodicity of the triple-step array is maintained with a precision of one interatomic distance in this surface region.

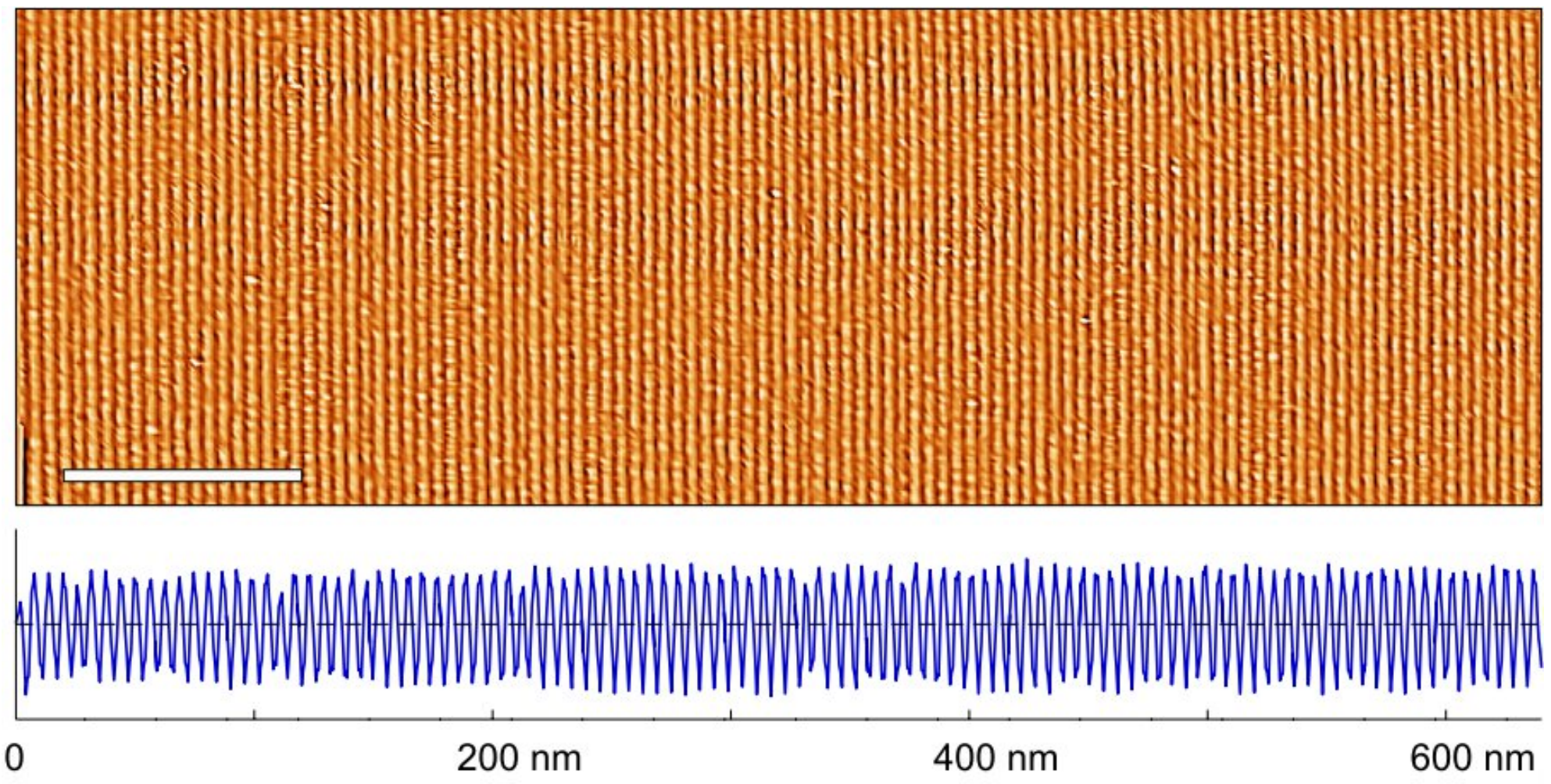

**Figure 2.** Typical differential *D(x,y)* map obtained from a large-area STM image of a regular triple-step array fabricated on the Si(5 5 7) wafer (top panel, image size of 640 × 210 nm$^2$, tunneling voltage $U$ = -1.8 V, tunneling current $I$ = 40 pA, scale bar corresponds to 100 nm). The solid line at the bottom shows the profile of the *D(x,y)* map averaged along the vertical direction and illustrates the regularity of the triple-step array on length scales exceeding 500 nm.

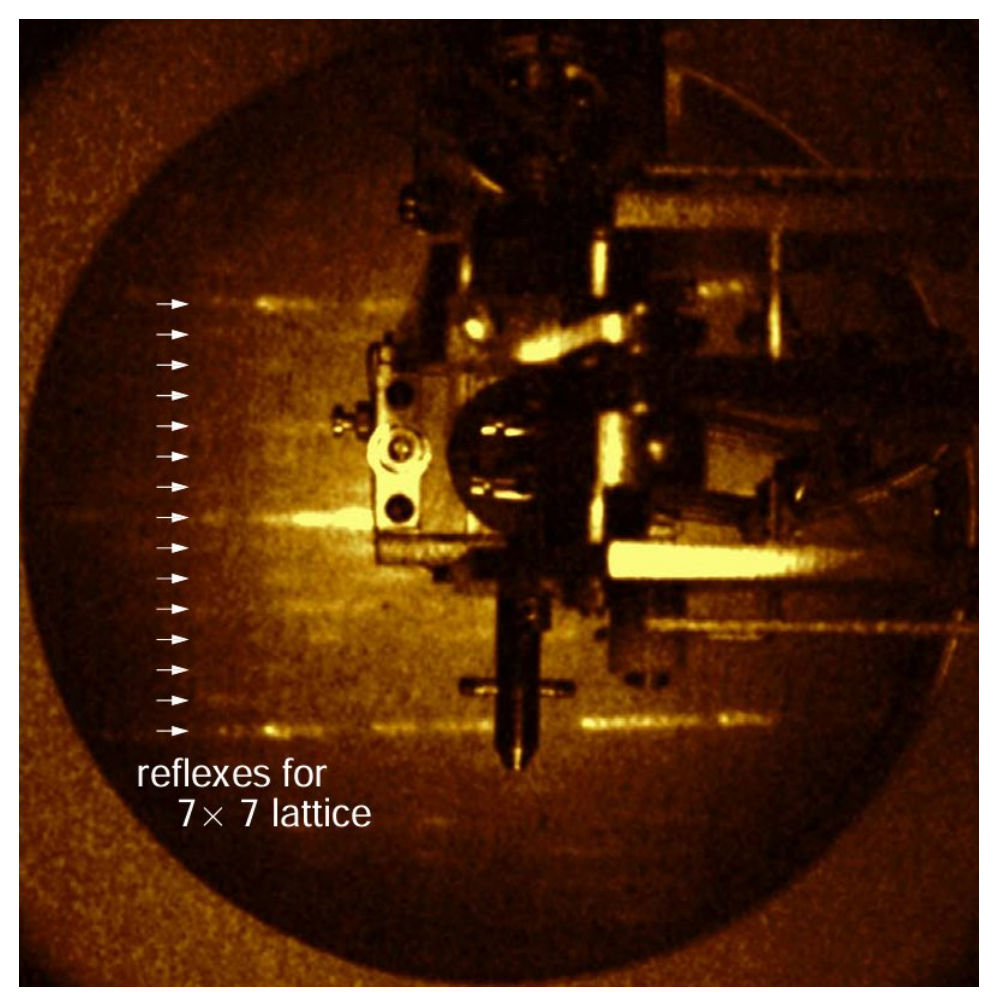

**Figure 3.** LEED pattern taken at an incident electron beam energy of 75 eV from the triple-step staircase fabricated on the Si(5 5 7) wafer. The marked horizontal rows correspond to the 7×7 reconstruction on the Si(1 1 1) terraces. The periodic spots in horizontal rows correspond to the formation of the regular triple-step array on large surface areas..

The formation of the periodic step array on large surface areas was also confirmed by LEED. As an example, the diffraction pattern shown in Fig. 3 demonstrates horizontal 7×7 rows due to formation of the 7×7 reconstruction on the terraces and sharp, regularly spaced spots within the rows due to the periodicity of the triple-step array. The equidistant spots along the horizontal rows prove the formation of the regular triple-step staircase with one preferential periodicity on large surface areas. To get reliable information about the step periodicity and possible atomic structure of the triple steps, a series of atomically resolved STM experiments were conducted on annealed Si(5 5 7) wafers providing high quality diffraction patterns. Since large area STM data can be distorted by undesirable drift artifacts and LEED studies have provided controversial conclusions [15, 16], only corrected to 7×7 and 5×5 unit cells *D(x,y)* maps obtained from atomically resolved topography images were analyzed.

**Results and discussion**

Figure 4a shows a typical atomically-resolved STM topography image of the periodic triple-step array formed on the Si(5 5 7) wafer demonstrating eight consecutive terraces and triple steps. As can be seen, the width of the terrace and step areas is comparable, corresponding to four rows of adatoms in the Si(1 1 1)-7×7 structure, or nine <110>-directed zigzag atomic rows in the 1×1 structure (9*b*). The terrace width in Fig. 4, which comprises a Si(1 1 1)-7×7 unit cell and an additional adatom row after the triple step, is consistent with values reported in several earlier works [1, 16]. Based on the dimensions of the Si(1 1 1)-7×7 reconstruction, the terrace width in this case equals $9b = 9 \times 0.3325$ nm = 2.99 nm.

To eliminate uncontrollable image distortions, the *D(x,y)* maps underwent affine correction to achieve ideal 7×7 unit cells on the terraces (Fig. 4c). After such correction, a precise determination of the distances between equivalent features of neighboring terraces can be done using either the *y*-averaged cross-sections (bottom panel of Fig. 4c) or the 7×7 lattice overlaid onto entire *D(x,y)* map. As an example, the model of the 7×7 reconstruction overlaid onto one of the terraces in the middle of the *D(x,y)* map in Fig. 4c illustrates that after the image correction, the positions of experimentally observed adatom features on the terraces and the positions of adatoms in the ideal Si(1 1 1)-7×7 structure coincide with a good accuracy (additional *D(x,y)* maps, where the distance between consecutive terraces can be determined in *b* units by comparison with the ideal 7×7 and 1×1 lattices overlaid onto entire maps in WSXM software, are shown in Figs. S1 and S2). The *y*-averaged cross-section of the *D(x,y)* map (bottom panel in Fig. 4c) reveals the distances between equivalent atomic features of neighboring terraces, defined with a precision of one interatomic distance (shown in units of *b* under the line profile). As indicated in Fig. 4c, the preferential periodicity *L* is equal to 18*b*, although the distances between consecutive triple steps corresponding to $L = 19b$ and $L = 17b$ are also observed. The STM data show that even in the case of a regular triple-step staircase formed on large surface areas, demonstrating sharp diffraction spots, the distance between consecutive triple steps can in some cases deviate by ±*b* from the mean and preferential step periodicity, which is close to 18*b*. This corresponds to neither the Si(5 5 7) [1] nor the Si(7 7 10) surface structure [16], which should provide a triple-step periodicity *L* close to 17*b* and 16*b*, respectively. The lateral distances obtained from the corrected *D(x,y)* maps in our case suggest the formation of a Si(8 8 11) triple-step staircase, which has not been discussed in the literature. The correction of the *D(x,y)* map is also confirmed by the fine structure of its *y*-averaged cross-section shown in Fig. 4c (bottom panel). It can be seen that for all “terrace + step” sequences corresponding to $L = 18b$ in Fig. 4c, the *y*-averaged profiles show four intense atomic features corresponding to four adatom rows on the terraces (7×7 unit cell plus an additional adatom row on the right side), as well as two intense features in the triple-step area. One of the triple step features can be related to adatoms resolved on the Si(1 1 1) mini-terrace after the monatomic step, while the other feature is related to the atomic reconstructions on the double step after the mini-terrace. For clarity, the features corresponding to the reconstructions on the Si(1 1 1) terraces and double steps are indicated in the cross-sections shown in Figure 5c. For wider triple steps corresponding to $L = 19b$, the cross-section of the *D(x,y)* map shows three intense peaks in the triple step region. Two of them, presumably, correspond to adatoms more clearly resolved on a slightly wider mini-terrace following the monatomic step, and the third one corresponds to the reconstruction on the double step after the mini-terrace. Note that the difference between the 18*b* and 19*b* sequences in the staircase is invisible in the typical plane-corrected (Fig. 4a) and aligned (Fig. 4b) topography images but is clearly distinguished in the *y*-averaged line profile of the differential *D(x,y)* map (bottom panel of Fig. 4c). Summarizing, we can conclude that this triple-step structure with four adatom rows on the terraces corresponds to the Si(8 8 11) staircase with $L = 18b \approx 5.99$ nm.

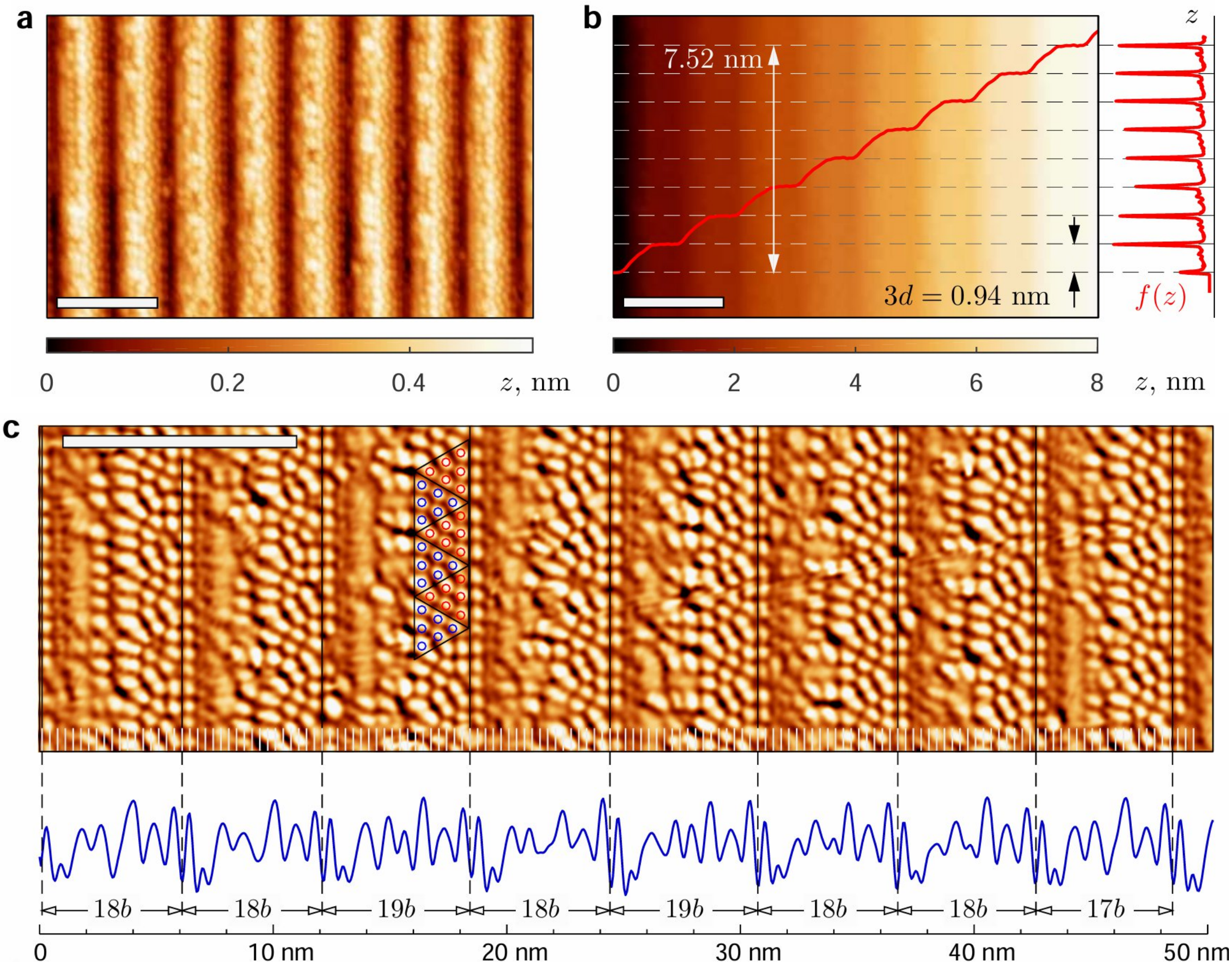

**Figure 4.** (a) STM topography image after global plane correction (image size 48.4 × 30 nm², *U* = 1.5 V, *I* = 60 pA) acquired at room temperature reveals a periodic array of triple steps and Si(1 1 1) terraces. Each terrace comprises one complete 7×7 unit cell and an additional row of adatoms. (b) After subtraction of a plane parallel to the terraces, the same image shows all flat (1 1 1) terraces horizontal. The red line shows the height variation of the aligned image as a function of the *x*-coordinate. The inset on the right side shows the histogram, illustrating the distribution of visible heights for the aligned image. The interval between the dashed lines equals $3d = 0.94$ nm. (c) Differential *D(x,y)* map derived from the STM image in panel (a) using the DOG approach, calibrated to the Si(1 1 1)-7×7 unit cell, and its *y*-averaged cross-sectional profile (bottom panel). The ideal 7×7 unit cells are overlaid onto one of the terraces for comparison with experimental features on the *D(x,y)* map. The distance between equivalent atomic features on the neighboring terraces in *b* units is shown under the profile. Scale bars in all panels correspond to 10 nm.

To visualize the complex structure of the steps and obtain reliable information on their heights and periodicity, we analyzed STM images after subtraction of a plane parallel to the terraces (Figs. 4b and 5a), as well as atomically-resolved differential *D(x,y)* maps corrected using the 7×7 reconstructions on the terraces (Figs. 4c and 5b). The cross-section shown by a red line in Fig. 4b demonstrates that the step height corresponds to three distances between neighboring Si(1 1 1) atomic layers (3*d* = 0.94 nm).

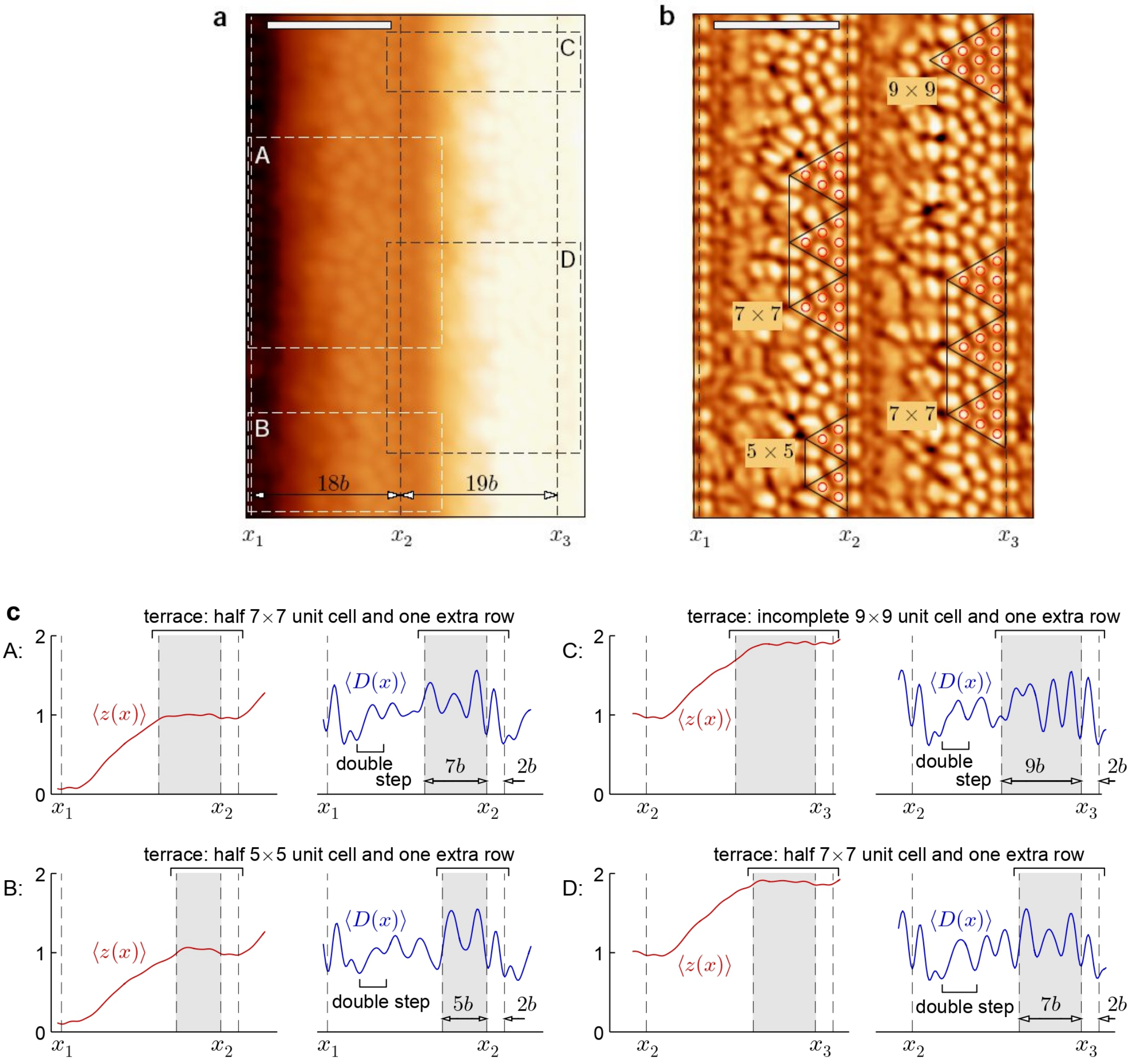

**Figure 5**. (a, b) Aligned topography STM image measured at tunneling voltage *U* = 1.5 V and current *I* = 60 pA (a) and differential *D(x,y)* map (b) of a surface area containing two periods of the triple-step structure shown in Figure 4. The distance between equivalent positions on the consecutive Si(1 1 1) terraces ($x_1$, $x_2$, $x_3$) is equal to 18*b* and 19*b*, as indicated on panel (a). The positions of adatoms corresponding to ideal 5×5, 7×7, and 9×9 cells are indicated above the experimental features on the *D(x,y)* map (b). Scale bars on panels (a) and (b) correspond to 5 nm. (c) *Y*-averaged cross-sections of the topography image and *D(x,y)* map in areas A-D indicated by rectangles on panel (a)The periodicity of the triple steps, the width of the terraces, the positions of the 5×5, 7×7, and 9×9 unit cells on the terraces and double steps after the mini-terraces between monatomic and double steps are indicated by vertical dashed lines and horizontal arrows on the profiles.

To illustrate the details of the atomic reconstructions on the terraces and triple steps, an STM image and a corresponding *D(x,y)* map of a smaller surface area are presented in Figure 5, demonstrating two “terrace + triple step” sequences with *L* = 18*b* and *L* = 19*b*. As can be seen from the differential map (Fig. 5b), in addition to the 7×7 reconstruction, which is dominant on the terraces of the triple-step staircase, fragments of the 5×5 and 9×9 reconstructions are also observed on the terraces (the positions of adatoms for ideal unit cells are shown above the *D(x,y)* map for surface areas A-D in Fig. 5b). The widths of the terraces containing 7×7 and 5×5 unit cells and additional row of adatoms after the triple step correspond to 9*b* and 7*b*, respectively, as illustrated on the *y*-averaged cross-sections of the areas A-D shown in Fig. 5c. As can be seen from Fig. 5c, the *y*-averaged line profiles of the $D(x,y)$ map in these regions reveal two, three, and four maxima within the unit cells (grey areas under the profiles), in accordance with the atomic structure of the 5×5, 7×7, and 9×9 reconstructions, respectively. Since the width of the "terrace + triple step" blocks in areas A and B is the same (18*b*), one can conclude that different terrace widths in these areas imply different triple step structures following the 9*b*- and 7*b*-wide Si(1 1 1) terraces.

The *y*-averaged cross-sections of the aligned topography image and the $D(x,y)$ map for areas *A*-*D* shown in Figure 5c demonstrate that in all cases the terrace is most probably followed by a triple step, which cannot be approximated by any particular, well-defined crystallographic plane (e.g., (1 1 2) or (1 1 3), as discussed earlier in the literature). Instead, the triple steps consist of a sequence of monatomic and double steps separated by a narrow Si(1 1 1) terrace. This breaking of the triple steps onto consecutive monatomic and double steps can be supposed from the structure of the y-averaged cross-sections of the aligned topography images and corresponding $D(x,y)$ maps (intense features corresponding to the double steps are marked under the profiles in Fig. 5c). The width of the mini-terrace differs for Si(1 1 1) terraces containing three rows of adatoms of the 5×5 reconstruction (area *B*, 7*b*-wide terrace) and four adatom rows of the 7×7 reconstruction (area *A*, 9*b*–wide terrace), which allows maintaining the same $L = 18b$ periodicity using different "terrace + triple step" units. Taking into account the structure of the *y*-averaged cross-section of the aligned topography image and the positions of the local minima in the *y*-averaged cross-section of the $D(x,y)$ map for areas *A* and *B* (left panel of Figs. 5c), two schematic models of the "terrace + triple step" units which can be responsible for the 18*b* periodicity on the Si(8 8 11) triple step staircase can be proposed, as illustrated in Figs. 6a and 6b. The simplified models of the Si(8 8 11) triple-step staircase shown in Fig. 6 take into account the widths of the terraces and positions of the monatomic and double steps in accordance with our STM data but do not include atomic reconstructions at the step edges (e.g., parallel and perpendicular dimers), defects on the terraces and triple steps [16, 17, 19], and breaking the periodic structure along the step edges frequently observed in our topography images and $D(x,y)$ maps (Figs. 4 and 5).

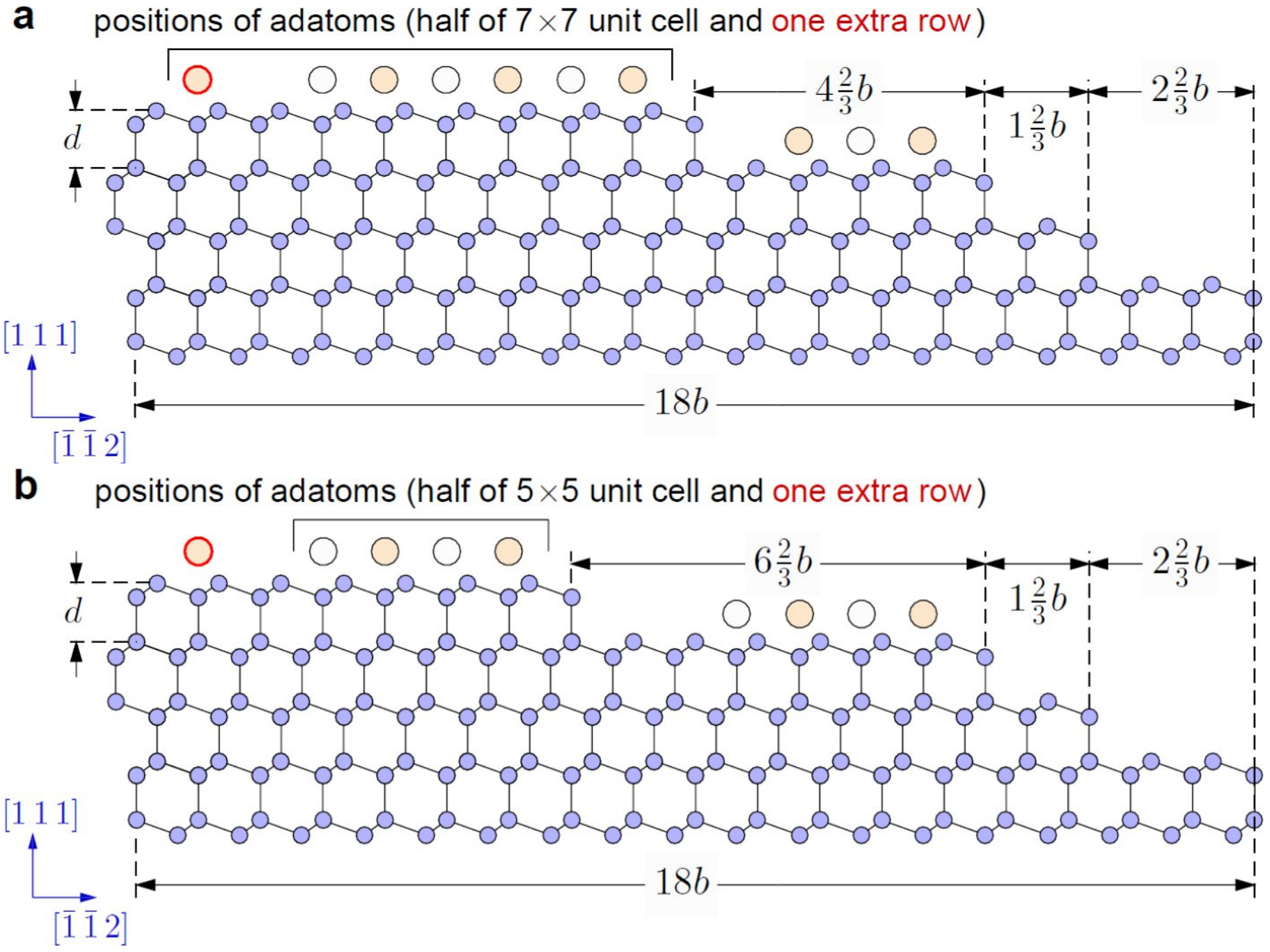

**Figure 6.** Schematic models of possible terrace and triple step structures responsible for the periodic Si(8 8 11) staircase in accordance with the STM images and *D*(*x*,*y*) maps shown in Figs. 4 and 5. The width of the top Si(1 1 1) terrace corresponds to half of the 7×7 and 5×5 unit cell with additional row of adatoms (shown by red circles) for panels (a) and (b), respectively.

Note that atomic features observed at the step edges in the STM images of the Si(*h h m*) stepped structures may not correspond to actual positions of silicon atoms at large positive bias voltages used for STM imaging of the Si(8 8 11) staircase (Figs. 4 and 5). These bias voltages are optimal for resolving adatoms on the Si(1 1 1) terraces but can not provide STM images with sharp steps and clearly resolved step edge features. As an example, Figures S3—S7 illustrate the bias voltage dependence of high resolution STM images obtained on Si(5 5 6) surface with smaller (5.05°) miscut angle and larger distance between the triple steps. In this case, scanning was possible even at small positive and negative bias voltages, which can hardly be achieved on the more corrugated Si(8 8 11) triple-step staircase because of tip breaking during scanning. Sharp monatomic step features and adatoms on the mini-terrace after the monatomic steps are well resolved only at small bias voltages below ±0.8 V (Fig. S3 and S7), while at higher bias voltages the step edge features are rather broad. Comparing the distances between the terrace and double step features in the averaged profiles of the *D*(*x*,*y*) maps in Fig. 5 and widths of the mini-terraces observed in the images measured at small bias voltages on the Si(5 5 6) surface (Figs. S3—S7), we can suggest that mini-terraces on the Si(8 8 11) triple-step staircase, most probably, contain two rows of adatoms (as shown in schematic models on Figs. 6a and 6b). The width of the double steps for the proposed models was estimated from the positions of the local minima on the *y*-averaged cross-sections of the *D*(*x*,*y*) map (indicated in Fig. 5c). The double step structure does not correspond to the local (1 1 2) [1, 17] or (0 0 1) orientations [16] but may correlate with the LEED simulations performed for the Si(2 2 3) and Si(5 5 6) triple step staircases in Ref. [21] and LEED experiments with spot profile analysis performed on Si(5 5 7) wafers by M. Henzler and R. Zhachuk [15].

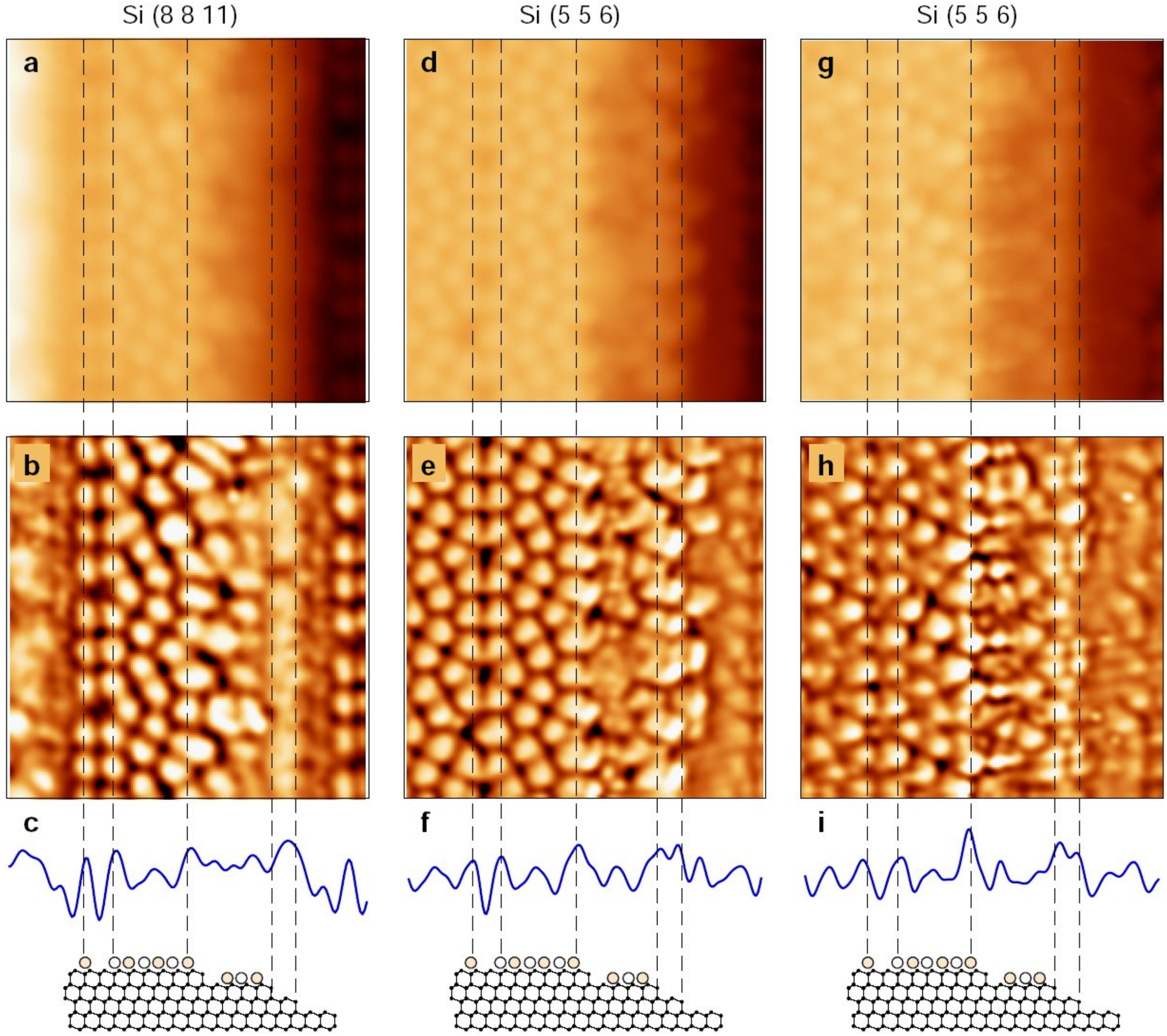

Figure 7. Comparison of the triple step STM images obtained for the Si(8 8 11) and Si(5 5 6) staircases with the model shown in Fig. 6a. (a-c) Aligned topography image of the triple step measured on the Si(8 8 11) staircase (a), differential map (b), *y*-averaged cross-section of the *D(x,y)* map (blue line) and schematic model resized in accordance with the dimensions of the scanned area (c). (d-i) Aligned topography images of the triple steps measured on the Si(5 5 6) staircase (d,g), differential map (e,h), *y*-averaged cross-sections of the corresponding *D(x,y)* maps (blue lines) and schematic model resized in accordance with the dimensions of the scanned areas (f,i). The STM images (82 × 82 Å$^2$) were measured at $U$ = 1.5 V and $I$ = 60 pA (a), $U$ = 0.6 V and $I$ = 50 pA (d), $U$ = -0.8 V and $I$ = 60 pA (g). Dashed vertical lines illustrate the coincidence of the positions of the experimental features with positions of adatoms and double step edges in the proposed model.

Comparison of the experimentally observed features with the triple step model shown in Fig. 6a is given in Figure 7. The schematic model is placed directly under the aligned topography images of the triple steps measured on the Si(8 8 11) and Si(5 5 6) staircases, *D(x,y)* maps and their *y*-averaged cross-sections (blue lines shown in panels c, f and i of Fig. 7). One can see that positions of adatoms on the terraces and double step features is in a good agreement with the proposed triple step model structure. For the images measured on the local Si(5 5 6) surface at $U$ = 0.6 V (Fig. 7d) and $U$ = -0.8 V (Fig. 7g), the double step features produce two distinct maxima in the *y*-averaged line profiles of the *D(x,y)* maps (Figs. 7f and 7i), which are not resolved for the Si(8 8 11) triple step staircase (Fig. 7c). Nevertheless, the positions of the local minima corresponding to the double steps are the same for all three line profiles and correlate well with the triple step model in Fig. 6a. Besides, the STM image and *D*(*x*,*y*) map shown in Figs. 7g and 7h, respectively, reveal parallel dimers and atomic reconstructions at the double step facet which are not seen in the Si(8 8 11) triple step images (Figs. 7a and 7b). Therefore, the schematic models of the triple steps proposed for the Si(8 8 11) staircase (Fig. 6) can be considered only as approximate and simplified because of step edge reconstructions and large number of defects both on the mini-terraces and at the step edges. Note that defects at the triple steps were typically observed in all SPM studies of the Si(*h h m*) staircases published previously.

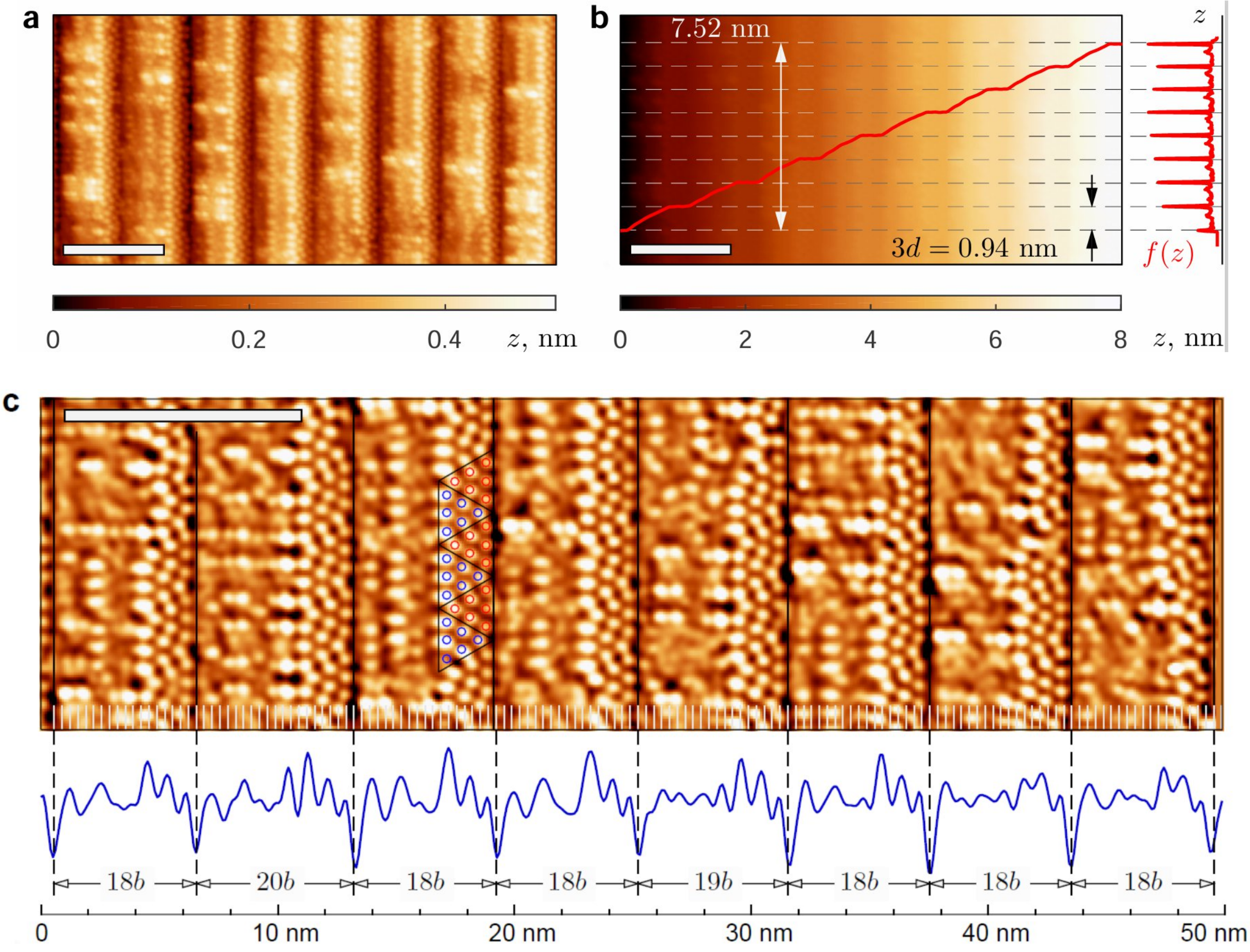

**Figure 8**. (a) A topography STM image of a periodic triple step system with terraces containing one unit cell of either 7×7 or 5×5 reconstruction without an additional row of adatoms. (b) The same STM image after subtracting a plane parallel to the terraces, and its *y*-averaged cross-section (red line overlaid on the image) illustrating triple step heights. Dashed horizontal lines show heights in multiples of $3d$ = 0.94 nm. The red line in the inset shows the histogram, illustrating the distribution of visible heights for the aligned image. (c) *D*(*x*,*y*) map obtained from the image (a) after correction using the ideal Si(1 1 1)-7×7 lattice parameters (top panel) and its *y*-averaged cross-section (bottom panel). The positions of the atoms of the ideal 7×7 structure are shown above the *D*(*x*,*y*) map for one of the terraces in the middle of the image. STM image (493×247Å$^2$) was measured in constant current mode ($U$ = 1.5 V, $I$ = 60 pA). Scale bars in all panels correspond to 10 nm.

Figure 8a shows atomically resolved STM image acquired in another surface area of the periodic triple-step staircase formed on the Si(5 5 7) wafer. As the differential *D*(*x*,*y*) map of the image (Fig. 8c) illustrates, the Si(1 1 1) terrace widths in this surface area corresponds exactly to half of either 7×7 (preferentially) or 5×5 (less frequently) unit cell without an additional row of adatoms on the terraces. The *y*-averaged cross-section of this image after subtracting the plane parallel to the Si(1 1 1) terraces (Fig. 8b) shows that height of the steps is equal to 3*d*, as was observed for the structure shown in Figures 4 and 5. As in the previous case, the *y*-averaged cross-sections of the topographic image (red line in Fig. 8b) and the corrected *D*(*x*,*y*) map (bottom panel in Fig. 8c) demonstrate the nonuniform structure of the triple steps, which consist of a consecutive monatomic step, a mini-terrace, and a double step, and cannot be approximated by a particular crystallographic plane. The cross-sections shown in Figs. 8b and 8c also suggest a preferential distance between consecutive triple steps $L = 18b$, which corresponds to the Si(8 8 11) orientation. As can be seen from the cross-section of the *D*(*x*,*y*) map (Fig. 8c), in some cases, the distance between consecutive terraces deviates by one (19*b*) or, very infrequently, two (20*b*) interatomic distances from the preferential triple step periodicity (18*b*), which corresponds to $L \approx 5.99$ nm. As for the *D*(*x*,*y*) map shown in Fig. 4c, the difference in step periodicity leads to a difference in the shape of the *y*-averaged line profile in the triple step area between the consecutive terraces (bottom panel of Fig. 8c). The correctness of the periodicity determination for the triple-step surface structure shown on Fig. 8 was additionally confirmed by suppression of the Fourier peaks in the 2D fast Fourier transform patterns obtained from the differential *D*(*x*,*y*) map [33].

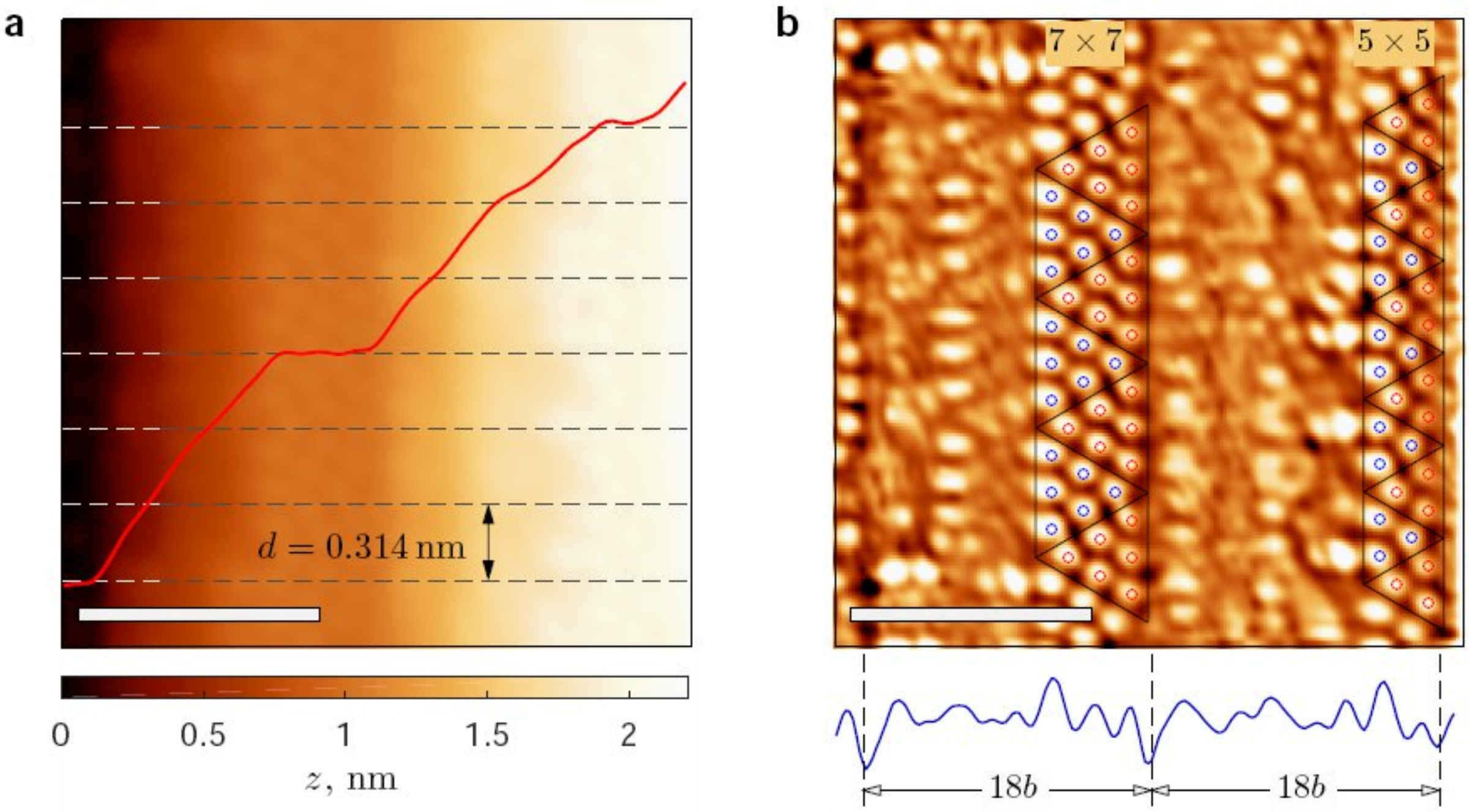

**Figure 9** (a) Topographic STM image of the triple-step structures with terraces containing half of the 7 × 7 (left) and 5 × 5 (right) unit cells without an additional row of adatoms. The red line shows the height variation of the aligned image as a function of the *x*-coordinate. (b) Differential *D*(*x*,*y*) map obtained from the STM image shown in panel (a) and its averaged cross-section (bottom panel). The positions of adatoms for the ideal 7 × 7 and 5 × 5 structures are overlaid onto the left and right terraces, respectively. The distance between neighboring dashed horizontal lines in (a) corresponds to one interplanar distance $d = 0.314$ nm. Scale bars in both panels correspond to 5 nm.

Figure 9 shows an STM image (Fig. 9a) of a structure with triple steps and terraces containing exactly half of either 7×7 (left part of the image) or 5×5 unit cell (right part of the image) and its differential *D*(*x*,*y*) map (Fig. 9b). The coincidence of the experimental features with the ideal 7×7 and 5×5 lattices overlaid onto the left and right terraces, respectively, proves the image correction. As can be seen from the averaged cross-sections (red line in Fig. 9a and bottom panel in Fig. 9b), both these atomic structures correspond to $L = 18b$, *i.e.*, to the Si(8 8 11) triple-step staircase. In the images shown in Figures 8 and 9, the 5×5-reconstructed Si(1 1 1) terraces coexist with the 7×7-reconstructed ones, without breaking periodicity or changing the step height.

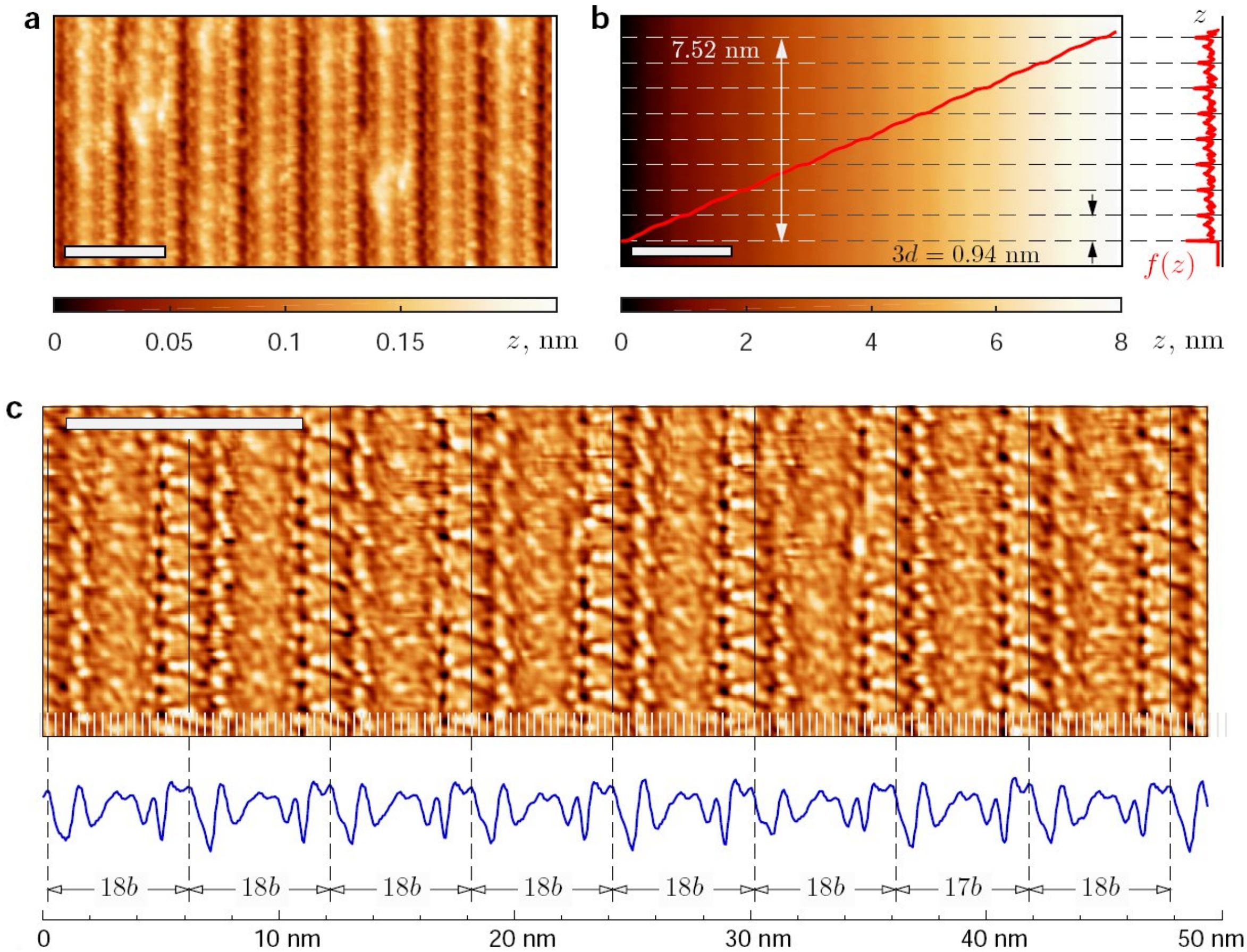

**Figure 10.** (a) A STM image of a periodic triple-step system with Si(1 1 1) terrace widths corresponding to half of the 5×5 unit cell. (b) Image shown in panel (a) after subtracting a plane parallel to the terraces, and its *y*-averaged cross-section (red line overlaid on the image) illustrating triple step height. Dashed horizontal lines in (b) show heights in multiples of $3d = 0.94$ nm. The red line in the inset (right panel) shows the histogram, illustrating the distribution of visible heights for the aligned image. (c) Differential $D(x,y)$ map obtained from the STM image (a) after correction to the Si(1 1 1)-5×5 lattice and its *y*-averaged cross-section (bottom panel). The period of this structure ($L = 18b \approx 5.99$ nm) corresponds to the Si(8 8 11) surface orientation. The STM image (50×25 nm$^2$) was measured in constant current mode at tunneling parameters $U = -1.4$ V and $I = 50$ pA. Scale bars in all panels correspond to 10 nm.

As our STM experiments reveal, the triple-step structure with Si(1 1 1)-5×5 terraces, corresponding to the Si(8 8 11) staircase, can be formed on rather large surface areas containing at least ten consecutive terraces and triple steps. Figure 10 shows an STM image and a $D(x,y)$ map of such a structure together with their *y*-averaged cross-sections. The topographic image in Fig. 10a reveals a triple-step system without periodicity breaking and clearly shows the non-uniform structure of the triple steps. At the large negative bias voltage ($U = -1.4$ V), the triple steps appear as sequences of a Si(1 1 1)-5×5 terrace, a monatomic step, another Si(1 1 1)-5×5 terrace, and a higher step with atomic reconstructions at the step facet. The plane-corrected STM image (Fig. 10b) and its *y*-averaged cross-section (the red line in Fig. 10b) suggest that the triple steps in this case consist of a consecutive monatomic step, a Si(1 1 1)-5×5 terrace, and a double step. The averaged profile of the affine-corrected *D(x,y)* map (bottom panel in Fig. 10c) confirms the formation of a stepped structure without periodicity breaking, with a period $L = 18b \approx 5.99$ nm, on this surface area.

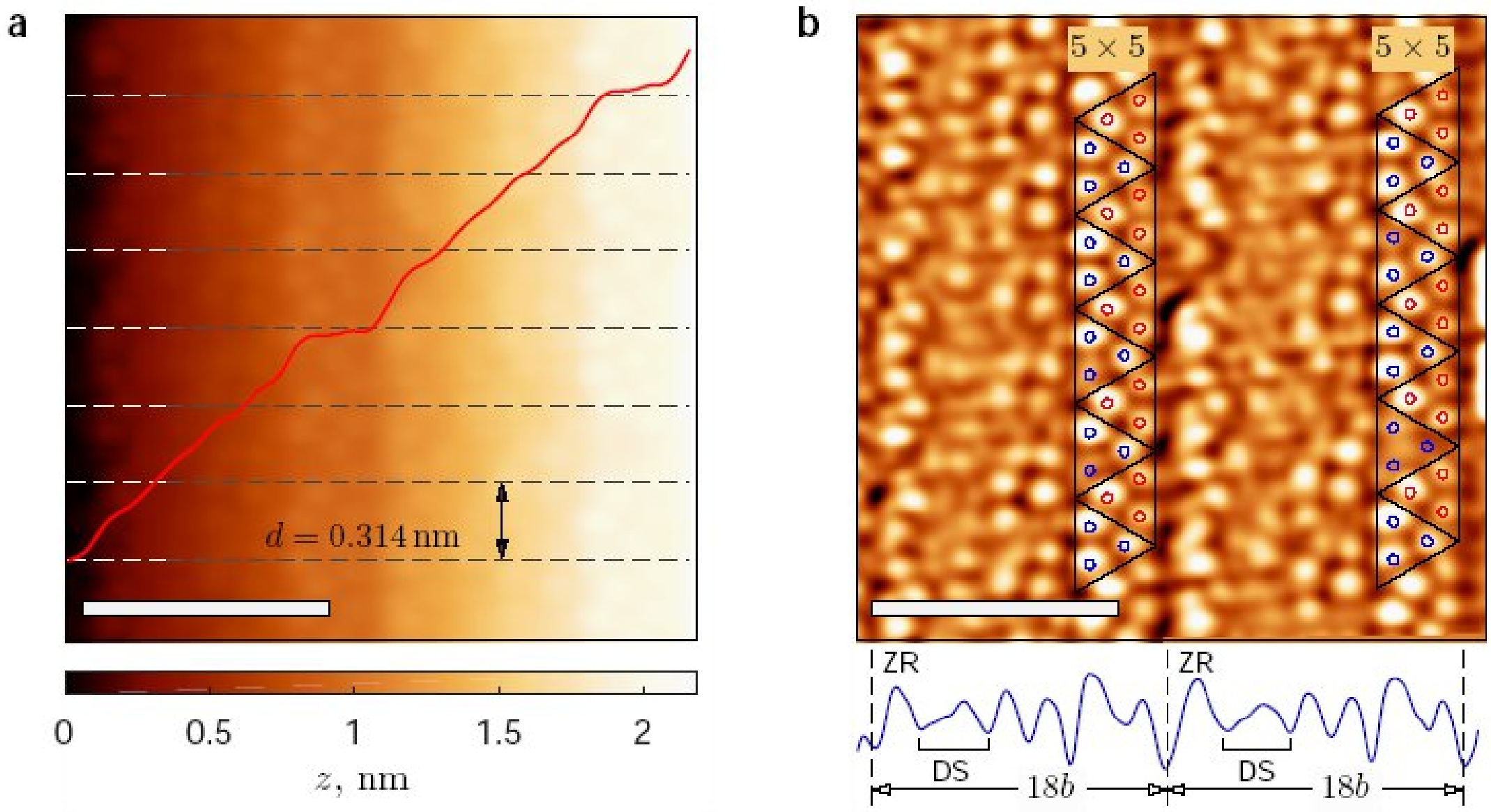

**Figure 11.** (a) STM image of the triple-step system with terraces containing one 5×5 unit cell after subtraction of a plane parallel to the Si(1 1 1) terraces. The red line overlaid on the image shows the height variation of the aligned image as a function of the *x*-coordinate. Dashed horizontal lines show heights in multiples of *d*=0.314 nm. (b) A differential *D(x,y)* map obtained from the STM image (a) after correction to the Si(1 1 1)-5×5 lattice (top panel) and its *y*-averaged cross-section (bottom panel). The ideal 5×5 lattices are shown above the terraces in the left and right parts of the *D(x,y)* map. STM image (12.6 × 12.6 nm$^2$) was measured at *U* = 1.4 V and *I* = 60 pA. Scale bars in both panels correspond to 10 nm. ZR and DS mean zigzag row and double step, respectively.

An unoccupied electron state STM image of such a triple-step structure (Fig. 11a) and its differential *D(x,y)* map (Fig. 11b) provide contrast images of the 5×5 reconstruction on the terraces but do not reveal the non-uniform structure of the triple step as a sequence of a monatomic step, a terrace, and a double step, which is observed in the filled-state images (Fig. 10a). The *y*-averaged cross-section of the *D(x,y)* map and comparison with the ideal 5×5 lattice shown in Fig. 11b indicate the formation of a triple-step structure with terraces containing half of the 5×5 unit cell, which also corresponds to the Si(8 8 11) surface orientation. Taking into account the experimental data presented in Figures 8–11, two alternative models of possible “terrace + triple step” units with terrace widths corresponding to the half of the 7×7 and 5×5 unit cells can be proposed for the Si(8 8 11) triple-step staircase, as schematically shown in Figs. 12a and 12b.

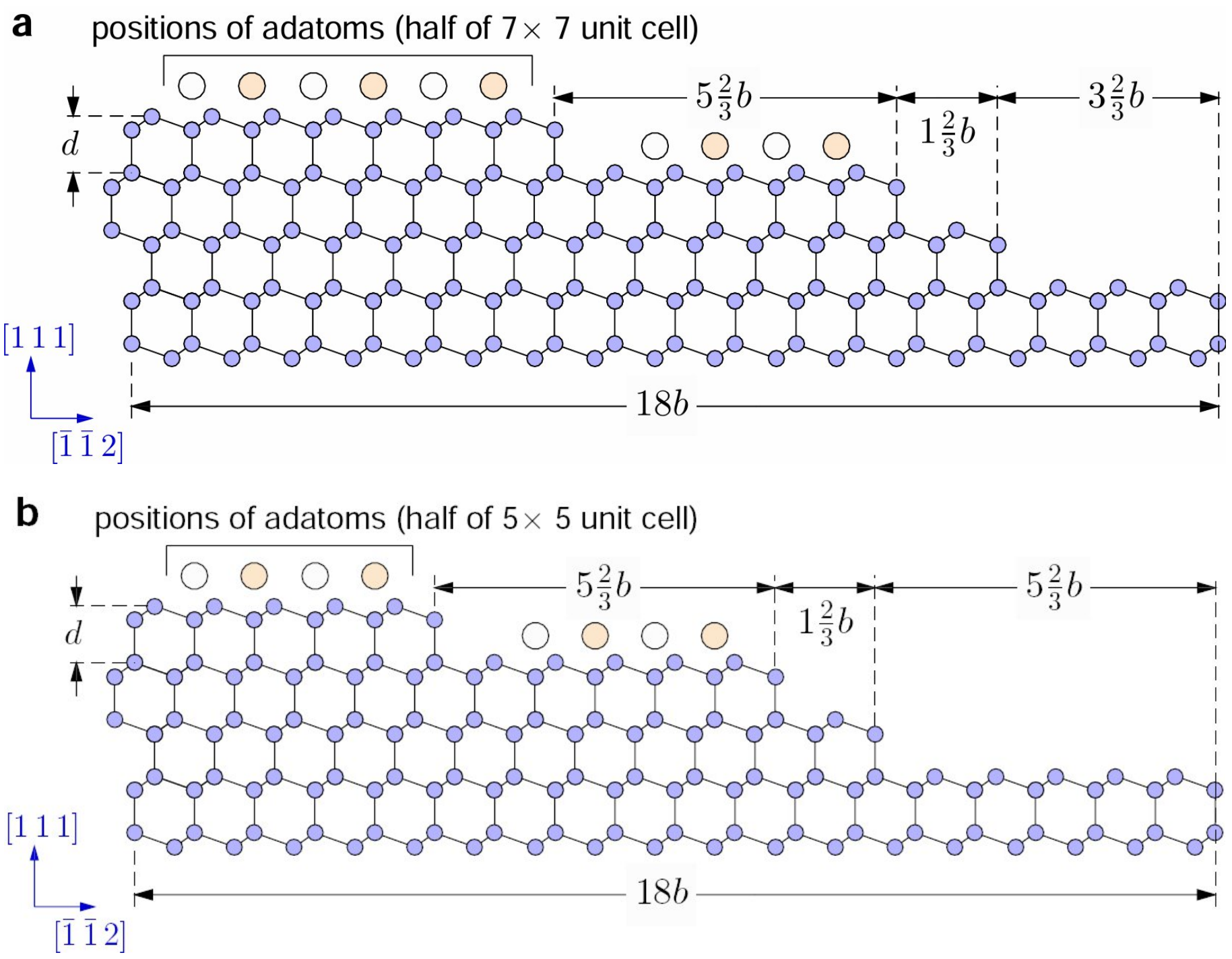

**Figure 12.** Schematic models of possible terrace and triple step structures responsible for the periodic Si(8 8 11) staircase in accordance with the STM images and *D(x,y)* maps shown in Figs. 8-11. The width of the top Si(1 1 1) terrace corresponds to half of the 7×7 (7*b*) and 5×5 unit cell (5*b*) for panels (a) and (b), respectively.

## Conclusion

The atomic structure of the periodic triple-step array fabricated on nominally Si(5 5 7) wafers has been studied using scanning tunneling microscopy. Detailed analysis of the experimental data indicates the formation of the Si(8 8 11) triple-step staircase with a period $L = 18b \approx 5.99$ nm, which seems to be in contrast with the Si(5 5 7) and Si(7 7 10) triple-step structures with periods of $17b \approx 5.65$ nm and $16b \approx 5.33$ nm, discussed earlier in the literature (here $b = 0.333$ nm is the distance between atomic rows for the Si(1 1 1)-1×1 surface). Atomically resolved STM data and differential maps obtained using the difference-of-Gaussians approach demonstrate that the distance between consecutive triple steps $L = (18 \pm 1)b$ can be maintained over rather large surface areas using different configurations of the triple steps and Si(1 1 1) terraces. The width of the terraces corresponds to the typical dimensions of the 5×5, 7×7, and 9×9 unit cells, while the triple steps consist of a sequence of monatomic and double steps separated by a narrow Si(1 1 1) terrace containing one or two Si adatom rows. Four schematic models of possible “terrace + triple step” atomic structures containing Si(1 1 1) terraces with widths corresponding to half of the 7×7 or 5×5 unit cell, with and without an additional adatom row, are presented. These models are in agreement with our STM data, but the possibility of other “terrace + triple step” units corresponding to $L = 18b$ is not excluded. The proposed triple step models do not consider possible atomic reconstructions at the step edges or defects produced by missing atoms, adsorbates, and doping atoms, which were beyond the scope of the current study. Nevertheless, the observation of several possible “terrace + triple step” configurations maintaining the same periodicity suggests that self-organization of the regular step array on large vicinal silicon surface areas may be related to step-step interaction and the appearance of adsorbates or doping atoms at the steps rather than to surface energy minimization due to a particular atomic reconstruction of the terraces and triple steps.

## Acknowledgments

This work was supported by the Russian State Contracts of the Institute for Physics of Microstructures RAS and the Osipyan Institute of Solid State Physics RAS and by the Grant from the Ministry of Science and Higher Education of the Russian Federation (no. 075-15-2024-632). The authors are grateful to Dr. Nikolay V. Abrosimov from Leibniz Institute for Crystal Growth (IKZ) for providing single-crystal silicon wafers.

**Supplementary**

Fig. S1. Differential *D(x,y)* map of the Si(8 8 11) triple step structure with terraces containing 4 adatom rows (half of the 7×7 unit cell + one adatom row) after affine correction. For precise determination of the step periodicity the model of the 7×7 structure is overlaid on the whole map. The positions of the experimental features coincide with the ideal structure for one of the terraces in the middle of the map.

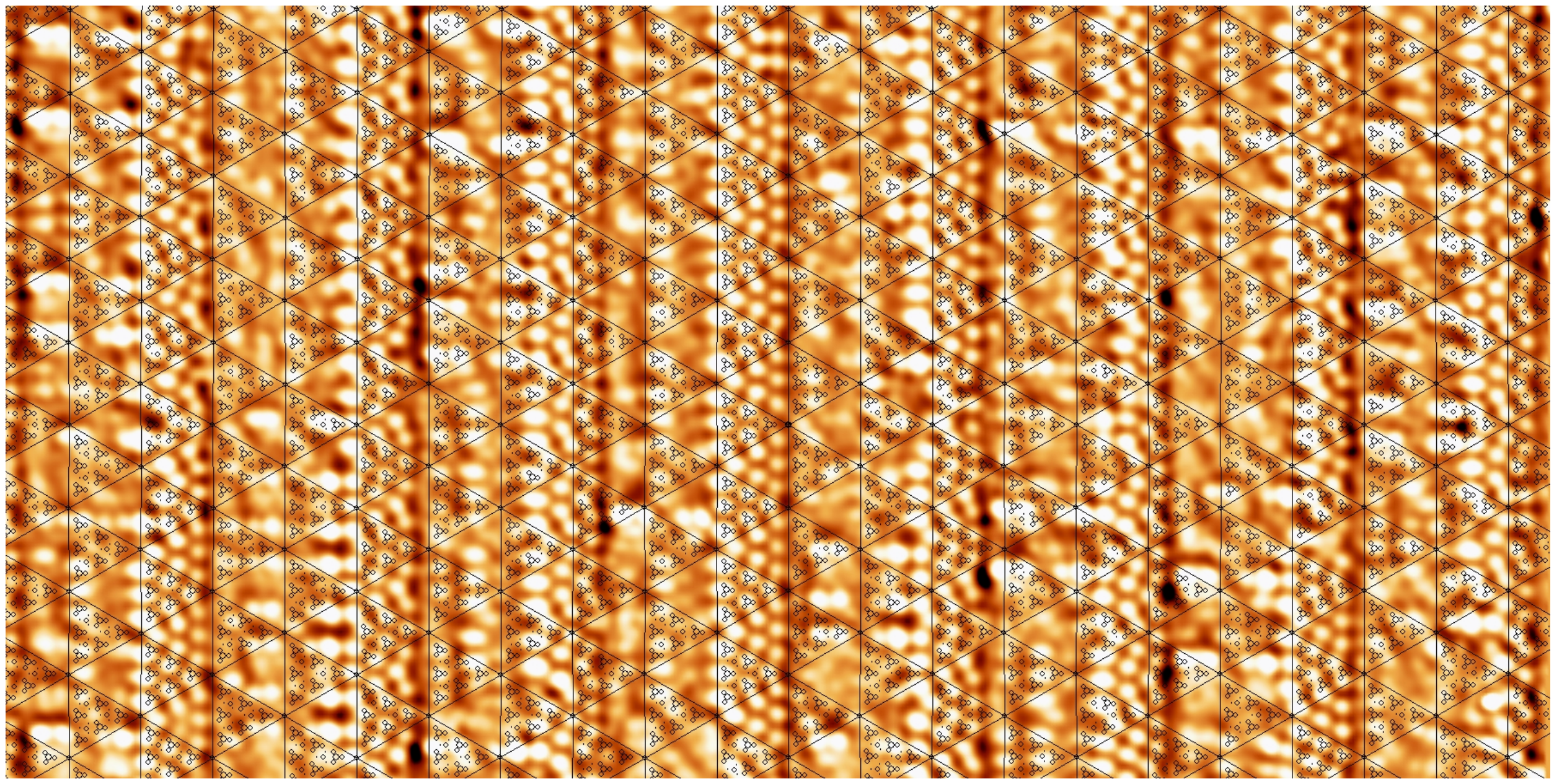

Fig. S2. Differential *D(x,y)* map of the Si(8 8 11) triple step structure with terraces containing 3 adatom rows (half of the 7×7 unit cell) after affine correction. For precise determination of the step periodicity the model of the 7×7 structure is overlaid on the whole map. The positions of the experimental features coincide with the ideal structure for one of the terraces in the middle of the map.

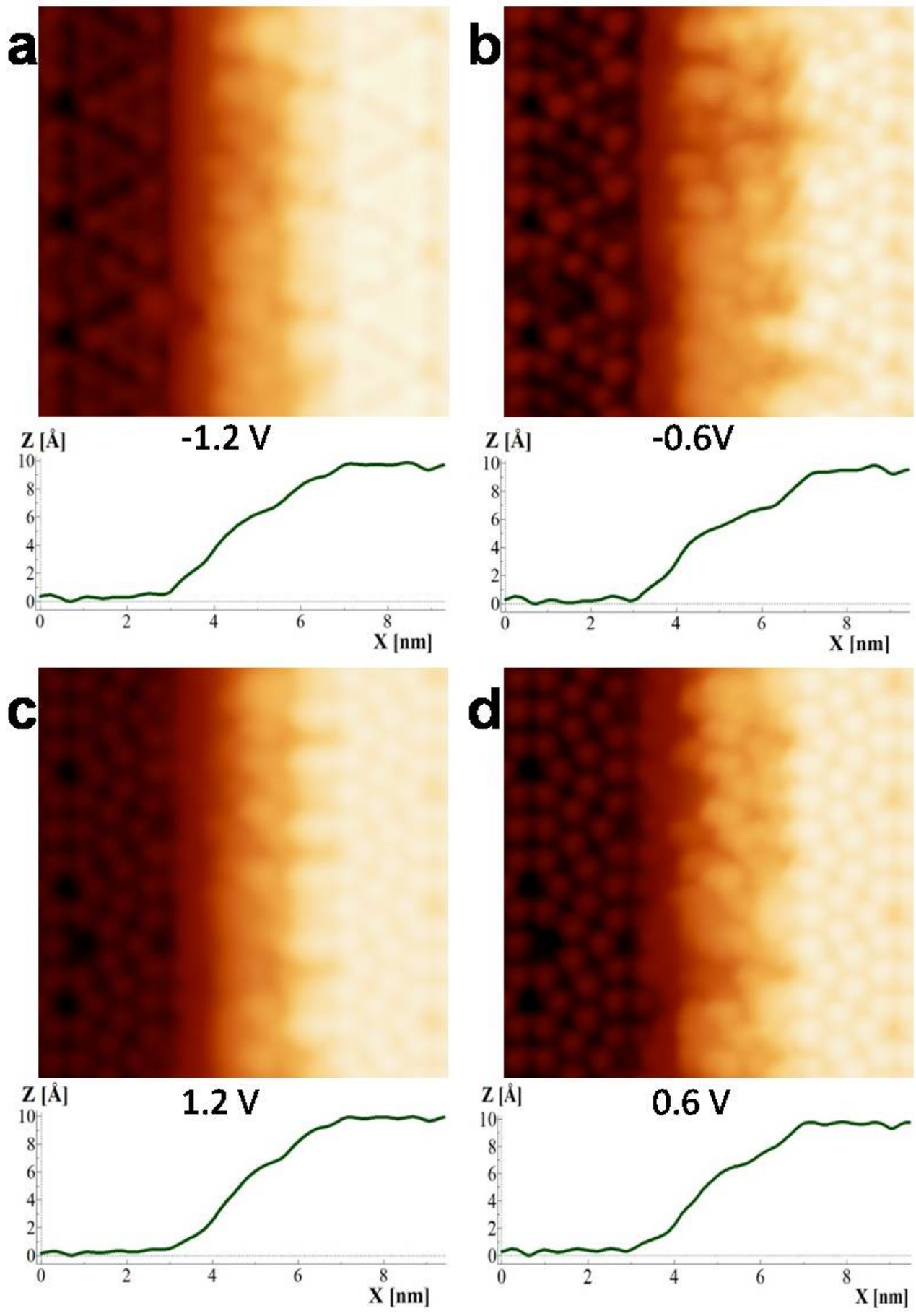

Figure S3 Bias voltage dependence of the STM image of triple step on a Si(5 5 6) staircase. *Y*-averaged cross-sections of the images are shown under each STM image to illustrate the difference in the triple step appearance at different voltages. The triple step is clearly resolved as a sequence of monatomic step, mini-terrace, and double step only at small voltages. Bright features resolved at $U = \pm 1.2$ V after the monatomic step correspond to the electron states at the step edges rather than adatoms on the mini-terrace.

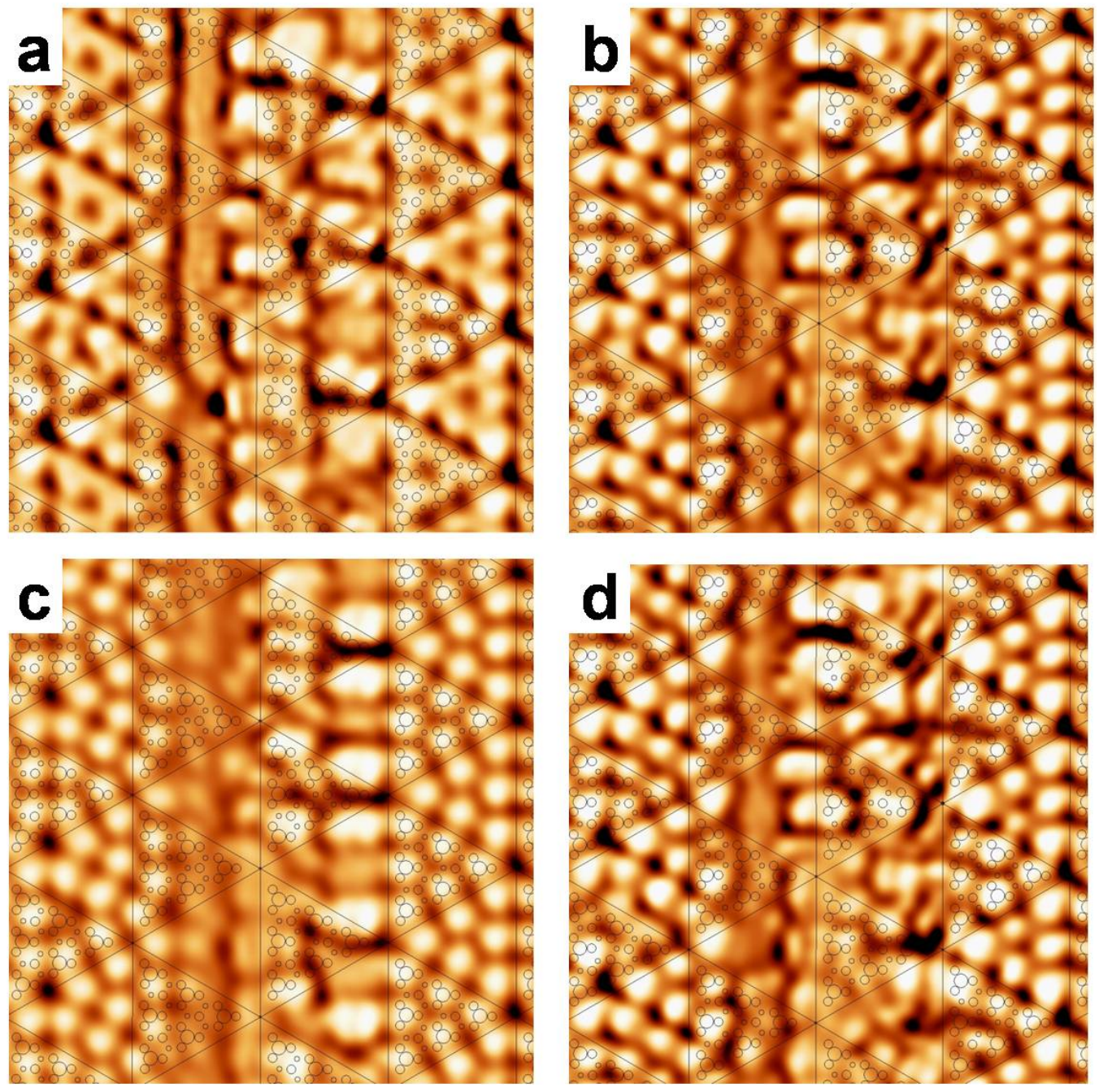

Figure S4. Differential *D(x,y)* maps obtained from the STM images on Fig. S3 and their *y*-averaged cross-sections. The 7×7 lattice overlaid onto the differential maps illustrate the coincidence of the experimental features with ideal 7×7 unit cell after the affine correction. The model lattices also show that the width of the triple and half of the 7×7 unit cell corresponds to 18*b*, similar to the STM images shown in Figures 7 and 8 and schematic model presented in Fig. 11a.

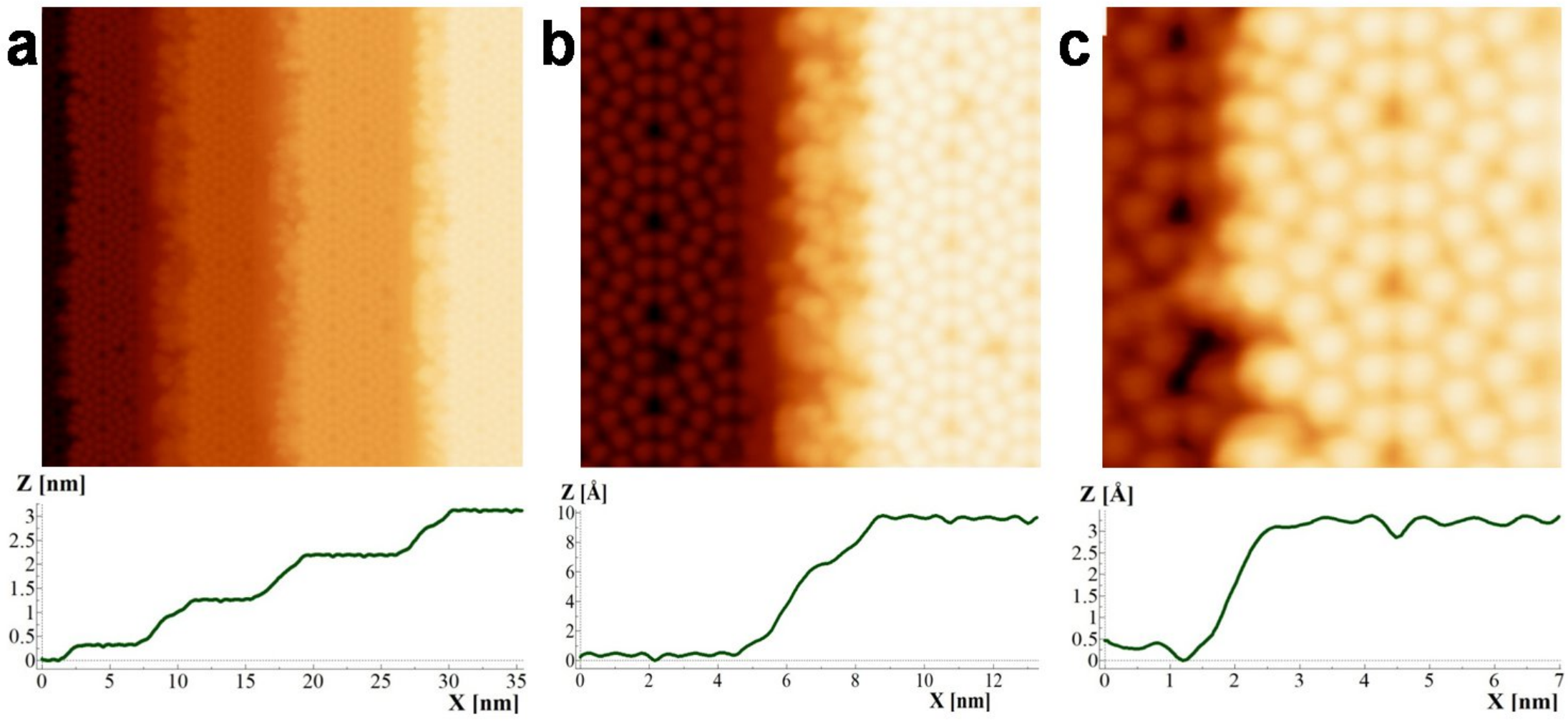

Fig. S5 (a) Aligned topography image measured at $U = 0.6$ V on a Si(5 5 6) wafer in a surface area containing triple and monatomic steps (top panel) and its *y*-averaged cross-section illustrating the step heights (bottom panel). (b) STM image of the triple step showing the triple step structure which can not be approximated by a particular plane (top) and its *y*-averaged cross-section (bottom). The image shows breaking the triple step onto consecutive monatomic step, mini-terrace with 7×7 reconstruction and double step. The monatomic step is well resolved and adatoms of the 7×7 reconstruction on the mini-terrace are clearly seen at this voltage. (c) STM image of the monatomic step from the left side of the surface area in panel (a). The image illustrates qualitatively the same structure of monatomic steps on panels (b) and (c).

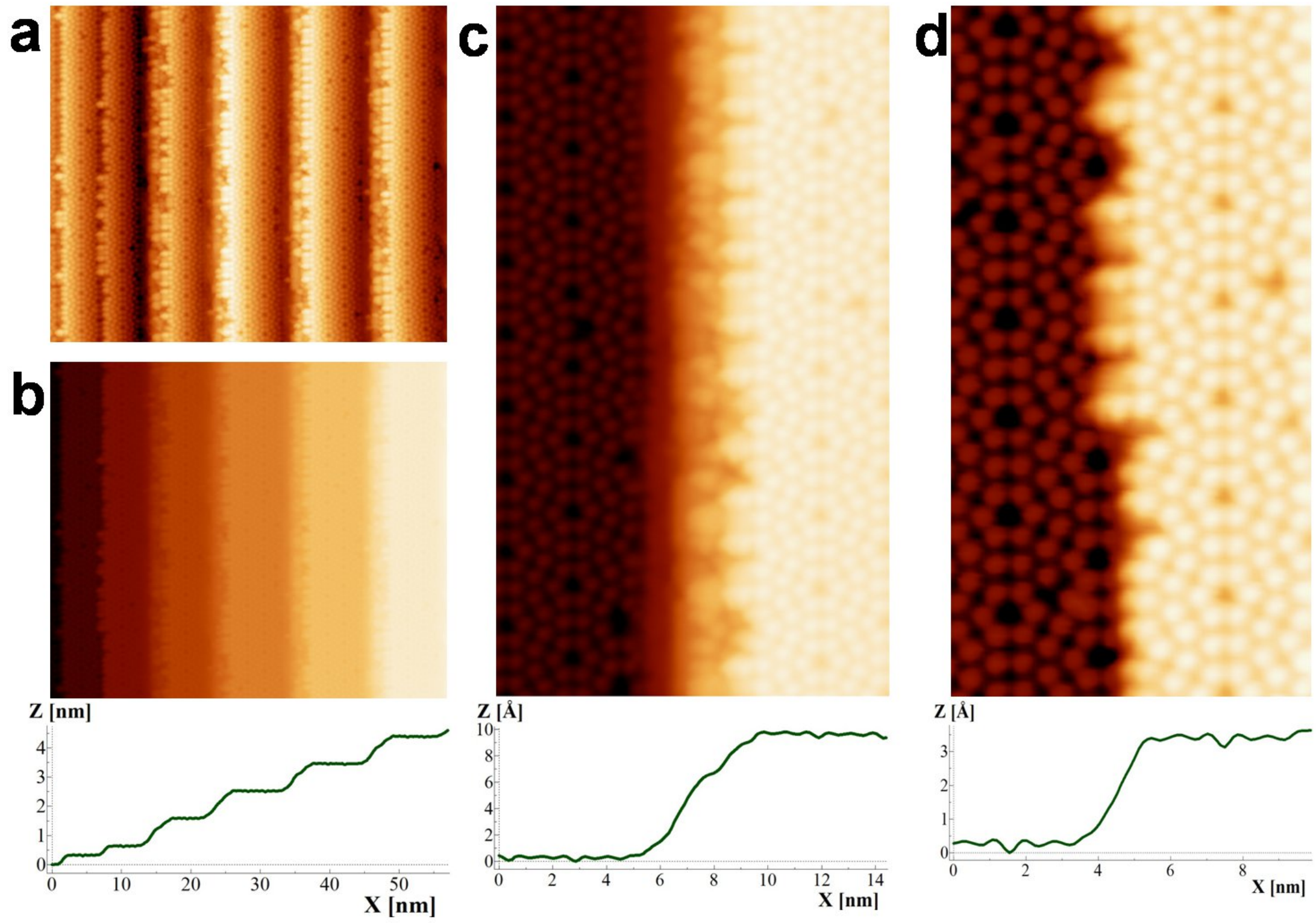

Fig. S6. (a) STM topography image measured at $U$ = 1.2 V on a Si(5 5 6) wafer in a surface area containing 4 triple and 2 monatomic steps. (b) The same STM image after subtraction of plane parallel to the terraces (top panel) and its averaged cross-section illustrating the step heights (bottom panel). (c,d) STM images of triple (c) and monatomic step (d) measured at U = 1.2 V and their averaged cross-sections. The appearance of monatomic step is the same when it followed by a wide Si(1 1 1) terrace with 7×7 reconstruction (d) and a mini-terrace of a triple step (c).

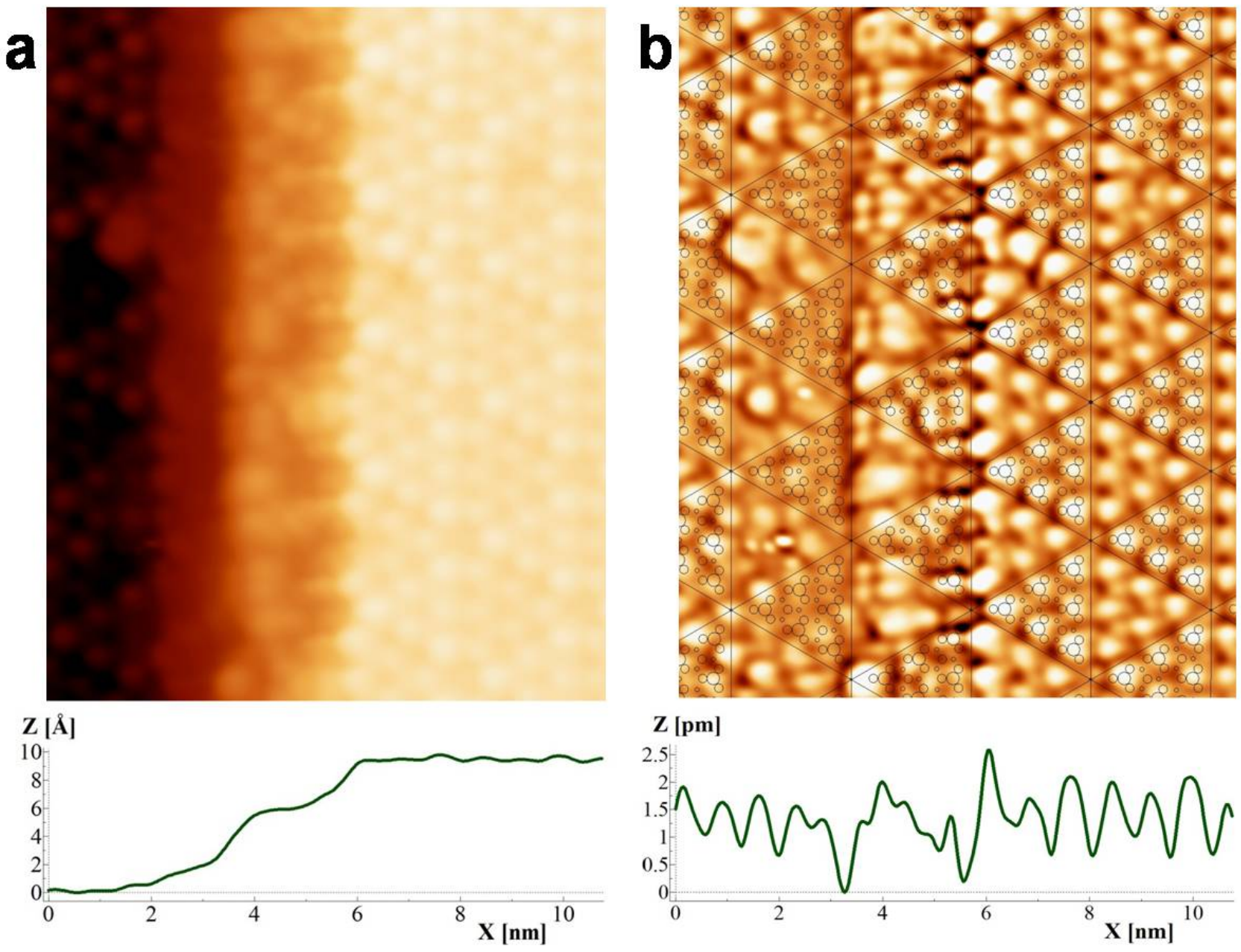

Fig. S7. High resolution STM image of the triple step on a Si(5 5 6) wafer measured at *U* = -0.8 V (a) and its differential *D(x,y)* map (b). The image clearly reveals consecutive monatomic and double steps as well as parallel and perpendicular dimers at the step edges. The 7×7 lattice overlaid onto the differential maps illustrate the coincidence of the experimental features with ideal 7×7 unit cell after the affine correction. The model lattices also show that the width of the triple step and half of the 7×7 unit cell corresponds to 18*b*, similar to the STM images shown in Figures 7 and 8.

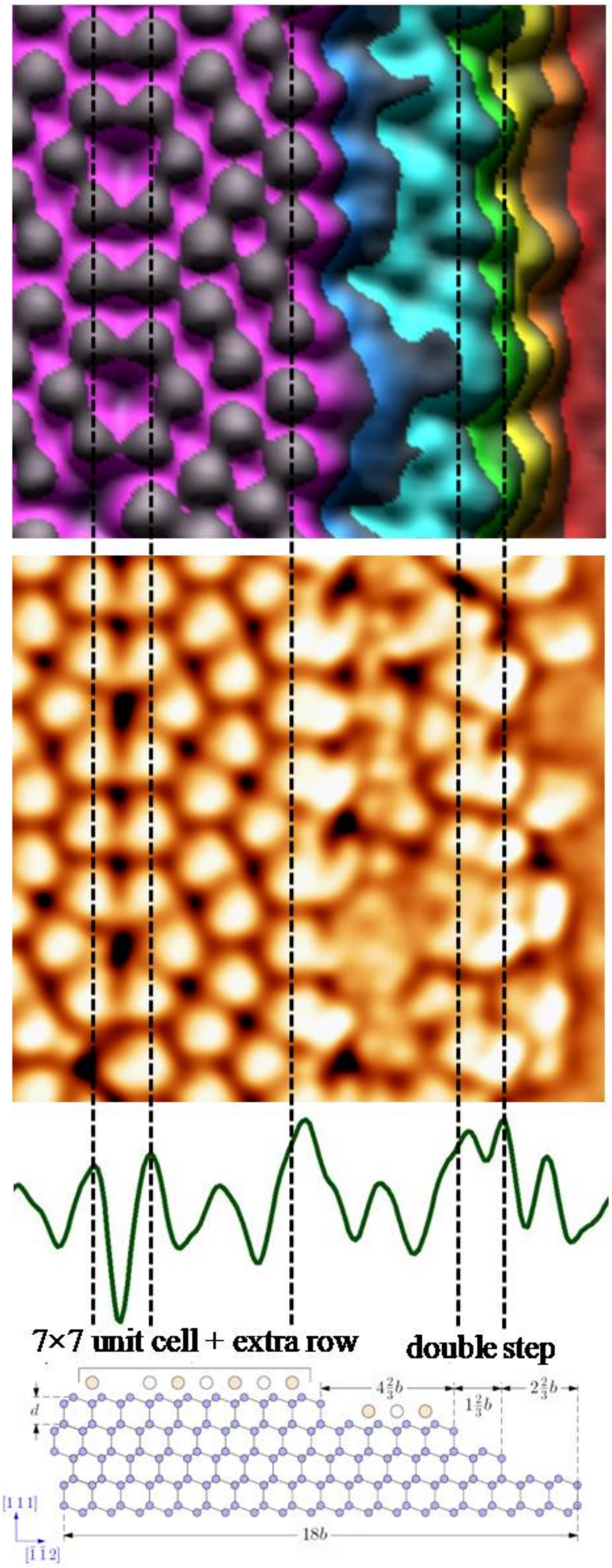

Fig. S8. Comparison of the experimental STM image measured on Si(556) surface and corresponding differential *D(x,y)* map with the triple step model shown in Fig. 6a. Top panel shows topographic image of the triple step in quasi-3D format. Middle panel demonstrates the *D(x,y)* map obtained from the STM image and its *y*-averaged cross-section. Dashed vertical lines illustrate the coincidence of the experimentally observed features with positions of adatoms on Si(111) terraces and double step edges in the schematic model (bottom panel).